\documentclass{conm-p-l}

\copyrightinfo{2016}{}

\usepackage{graphicx}
\usepackage{subfigure}

\usepackage{amssymb,amsmath,amsthm,amscd}

\newtheorem{theorem}{Theorem}[section]

\theoremstyle{definition}
\newtheorem{definition}[theorem]{Definition}

\theoremstyle{remark}

\numberwithin{equation}{section}

\newcommand{\ga}{\gamma}

\newcommand{\dl}{\delta}

\newcommand{\al}{\alpha}
\newcommand{\be}{\beta}

\newcommand{\pa}{\partial}

\newcommand{\la}{\lambda}

\newcommand{\om}{\omega}
\newcommand{\Om}{\Omega}

\newcommand{\non}{\nonumber}

\begin{document}
   
\title[A mathematical model of demand-supply dynamics]{A mathematical model of demand-supply dynamics with collectability and saturation factors}

\author{Y. Charles Li}
\address{Department of Mathematics, University of Missouri, 
Columbia, MO 65211, USA}
\email{liyan@missouri.edu}
\urladdr{http://faculty.missouri.edu/~liyan}

\author{Hong Yang}
\address{Mathematics of Networks and Communications Research Department,
Bell Laboratories, 600 Mountain Avenue, Murray Hill, NJ 07974, USA}
\email{h.yang@research.bell-labs.com}

\curraddr{}
\thanks{}

\subjclass{Primary 91, 34}

\date{}

\dedicatory{}


\begin{abstract}
We introduce a mathematical model on the dynamics of demand and supply incorporating collectability and saturation factors. Our analysis shows that when the fluctuation of the determinants of demand and supply is strong enough, there is chaos 
in the demand-supply dynamics. Our numerical simulation shows that such a chaos is not an attractor (i.e. 
dynamics is not approaching the chaos), instead a periodic attractor (of period 3 under the Poincar\'e period map) exists near the chaos, and co-exists with another periodic attractor (of period 1 under the Poincar\'e period map) near the market equilibrium. Outside the basins of attraction of the two periodic 
attractors, the dynamics approaches infinity indicating market irrational exuberance or flash crash. The period 3
attractor represents the product's market cycle of growth and recession, while period 1 
attractor near the market equilibrium represents the regular fluctuation of the product's market. Thus our model
captures more market phenomena besides Marshall's market equilibrium. When the fluctuation of the determinants of demand and supply is strong enough, a three leaf danger zone exists where the basins of attraction of all 
attractors intertwine and fractal basin boundaries are formed. Small perturbations in the danger zone can lead
to very different attractors. That is, small perturbations in the danger zone can cause the market to experience
oscillation near market equilibrium, large growth and recession cycle, and irrational exuberance or flash 
crash. 
\end{abstract}

\maketitle
\tableofcontents

\section{Introduction}

The dynamics of demand and supply is the key for a market. One can observe the demand and supply dynamics in action from common commodities like houses \cite{DLS14} and gasoline \cite{AEG13} \cite{Pla14}. Due to the huge surplus in supply, gasoline price has sharply dropped recently \cite{Pla14}. Gasoline price and housing 
price affect people's daily life. It is paramount to build better mathematical models on demand and supply dynamics. Various mathematical tools have been developed in studying demand and supply (see e.g. 
\cite{Heo14} \cite{TX14} \cite{Wei07}). Here we are employing dynamical system tools to study the dynamics 
of demand and supply generalizing the classical Marshall model. 

Our model shall describe the demand-supply dynamics of the global market on one product. This is the topic of Microeconomics \cite{Sto12} \cite{Oln09}. We 
are not dealing with aggregate demand and aggregate supply which belong to Macroeconomics \cite{KMS14}. The one product global market that we are 
modeling is close to a competitive market where the market dynamics is more objective in contrast to monopolistic, oligopolistic, and monopolistic competitive
markets. In an ideal competitive market, no individual buyer or seller can influence the price, the product feature is standardized, buyers and sellers are well informed, 
and firms can enter and leave with no significant barrier \cite{Sto12}. The law of demand states that as the price increases, the quantity demanded decreases, 
{\em ceteris paribus} (i.e. holding all other factors constant). On the other hand, the law of supply states that as the price increases, the supply quantity increases, 
{\em ceteris paribus}. According to Alfred Marshall, the demand curve and supply curve intersect at a market equilibrium (Fig. \ref{dse}). The market dynamics 
approaches the equilibrium, a phenomenon that Adam Smith called an``invisible hand" leading the market dynamics to the equilibrium. This is the classical theory 
on demand-supply dynamics. According to this theory, an individual firm in a competitive market is a price taker (the price of its product is set by the market equilibrium). 
Such a demand-supply model is very ideal. In reality, other factors (that are held constant in the statements of the laws of demand and supply) 
change dramatically and have significant effect on demand-supply dynamics. The prices of most products are not staying close to their equilibrium values. For example, 
the watch market is quite close to a competitive market. But individual firms are not price takers. Rolex watch price is much higher that those of less known brand watches. 
Even the average price of watches change substantially in time. Those factors that are held constant in the statements of the laws of demand and supply are called 
determinants \cite{Sto12}. The main determinants of demand are \cite{Sto12}:
\begin{enumerate}
\item taste and preference,
\item income level,
\item prices of related goods,
\item the number of potential buyers,
\item future expectation on the product and income.
\end{enumerate}
The main determinants of supply are \cite{Sto12}:
\begin{enumerate}
\item production technology,
\item costs of resources,
\item prices of other commodities,
\item future expectations on the product,
\item the number of potential sellers,
\item taxes and subsidies.
\end{enumerate}
In the statements of the laws of demand and supply, the amounts of demand and supply are amounts during a time period. The  demand and supply curves do not 
depend on time, and they are static curves. In our demand-supply model, the amounts of  demand and supply depend on time (they are the amounts at that time, not 
during a time period), and the price also depends on time. The demand-supply dynamics is represented by the temporal evolution of the price and the amounts of  
demand and supply. During this evolution, the determinants of  demand and supply constantly influence the dynamics. Such models are closer to the reality. Due to 
the variation of price from firm to firm, the price of the product's entire market is defined to be the average price.
\begin{definition}
The price $P(t)$ of a product at time $t$ is defined to be the average price over the product's global market at time $t$. The amount of demand $D(t)$ is defined to 
be the total amount of demand in the product's global market. The amount of supply $S(t)$ is defined to be the total amount of supply in the product's global market.
\end{definition}
Our model will be represented by a system of ordinary differential equations involving $P(t)$, $D(t)$ and $S(t)$. Earlier studies on differential equation models 
\cite{Kob96} \cite{SBD12} focused on the following equation
\[
\frac{dP}{dt} = f(D(P)-S(P))
\]
where time delay may be involved. Convergence of the dynamics to market equilibrium was the main interest \cite{Kob96} \cite{SBD12}. Our model takes  the general form 
\begin{eqnarray}
&& \frac{dP}{dt} = f_p(D-S), \non \\
&& \frac{dD}{dt} = f_d (P_d - P) + F_d(t), \label{gm} \\
&& \frac{dS}{dt} = -f_s(P_s-P) +g_s(D-S) +F_s(t),  \non
\end{eqnarray}
where ($f_p,f_d,f_s,g_s,F_d,F_s$) are general functions for now, and $P_d$ and $P_s$ are threshold prices of demand and supply. The price equation states that price change is 
determined solely by $D-S$. Price increases when $D-S>0$, and decreases when $D-S<0$. The change in the amount of demand $D$ depends on the price 
relative to the threshold price of demand. Usually the difference $D-S$ has little influence on buyers, and buyers just buy whenever they need and can afford the product. 
So $\frac{dD}{dt}$ has little dependence on $D-S$. On the other hand, $D-S$ has more significant influence on producers, and producers neeed to know $D-S$ to project the future 
trend of price and profit. So $\frac{dS}{dt}$ depends on $D-S$ next to $P_s-P$ (the functions $g_s$ and $f_s$). The determinants of demand also influence $\frac{dD}{dt}$.
Taste and preference can change with time. They can also change with price. When the price gets very high, the product may turn into a collectable product  $f_d >0$ (collectability factor). 
When the price gets extremely low, the market is over saturated, and the product may become less desirable $f_d<0$ (saturation factor). The income level and the prices of 
related goods can fluctuate in time, and they can be represented by a function of time $F_d(t)$ that is independent of the three variables ($P,D,S$). The determinants of supply 
also influence $\frac{dS}{dt}$. Production technology, costs of resources, and price of other commodities, and taxes and subsidies change with time, and they can be represented 
by a function of time $F_s(t)$ that is independent of the three variables ($P,D,S$). 

Collectability and saturation factors can also be used to model certain stocks. Some stock's price 
may get much higher than its true value, and more and more investors continue to buy them. On the other hand, when the stock's price gets much lower than its true value, still less and less investors want 
to buy them.

Next we set up a simple specific model of the demand-supply dynamics starting from (\ref{gm}). In general, we expect ($f_p,f_d,f_s,g_s$) 
to be linear only near zero, but for $f_p$ we believe that linear approximation should perform very well based on the general principle that price increases (decreases) when demand is 
more (less) than supply. We choose
\[
f_p (D-S) = \al (D-S)
\]
where $\al >0$ is a parameter. That is, the change rate in price is proportional to $D-S$. We choose
\[
f_d(P_d-P) = \be (P_d-P) [1-\be_1 (P_d-P)^2],
\]
where $\be >0$ and $\be_1>0$ are parameters, and we take into account collectability and saturation factors. When the price is too low, the market is already saturated, and the demand 
will not increase anymore. Since the lowest price is zero, we have 
\begin{equation}
1< \be_1 P_d^2 . \label{sc}
\end{equation}
When the price gets too high, the product may become a collectable item, and the demand can increase. For the supply equation, we choose
\[
f_s(P_s -P) = \ga (P_s -P), \  g_s (D-S) = \dl (D-S),
\]
where $\ga >0$ and $\dl >0$ are parameters. High price and high demand are positive factors for supply increase. In summary, we arrive at the following simple specific model for the 
demand-supply dynamics incorporating collectability and saturation factors:
\begin{eqnarray}
&& \frac{dP}{dt} = \al (D-S), \non \\
&& \frac{dD}{dt} = \be (P_d-P)[1-\be_1 (P_d-P)^2] + F_d (t), \label{sm} \\
&& \frac{dS}{dt} =-\ga (P_s-P) +\dl (D-S) +F_s(t), \non
\end{eqnarray}
where again ($\al , \be , \be_1 , \ga , \dl , P_d,P_s$) are positive parameters. For the functions $F_d(t)$ and $F_s(t)$, we can choose 
\begin{eqnarray}
&& F_d(t) = a \sin (\om_1t) , \label{df} \\ 
&& F_s(t) = c+ b \sin (\om_2t), \label{sf} 
\end{eqnarray}
where the income level and prices of related goods often fluctuate, and these factors lead to the oscillatory nature of $F_d(t)$. Production technology is represented by the constant term 
$c$ in $F_s(t)$. Costs of resources, prices of other commodities, and taxes and subsidies often fluctuate, and they are represented by the oscillatory sine term in $F_s(t)$. 

In the third equation of (\ref{sm}), 
\[
P_s -P = P_d -P +(P_s-P_d),
\]
and the ($P_s-P_d$) term can be incorporated into the $c$ term in $F_s(t)$. We will set $c=b=0$, and 
$F_d(t)$ term is enough to represent the fluctuation factor of determinants of supply and demand. Let 
\[
p=P-P_d, \  q =D-S,
\]
where $p \geq - P_d$, then the system (\ref{sm})-{\ref{sf}) takes the form
\begin{eqnarray}
&& \frac{dp}{dt} = \al q , \label{am1} \\
&& \frac{dq}{dt} = -\be p (1-\be_1 p^2) - \ga p - \dl q +a \sin (\om_1t), \label{am2}
\end{eqnarray}
where ($p,q$) = ($0,0$) represents the market equilibrium.

\begin{figure}[ht]
\centering
\includegraphics[width=4.5in,height=3in]{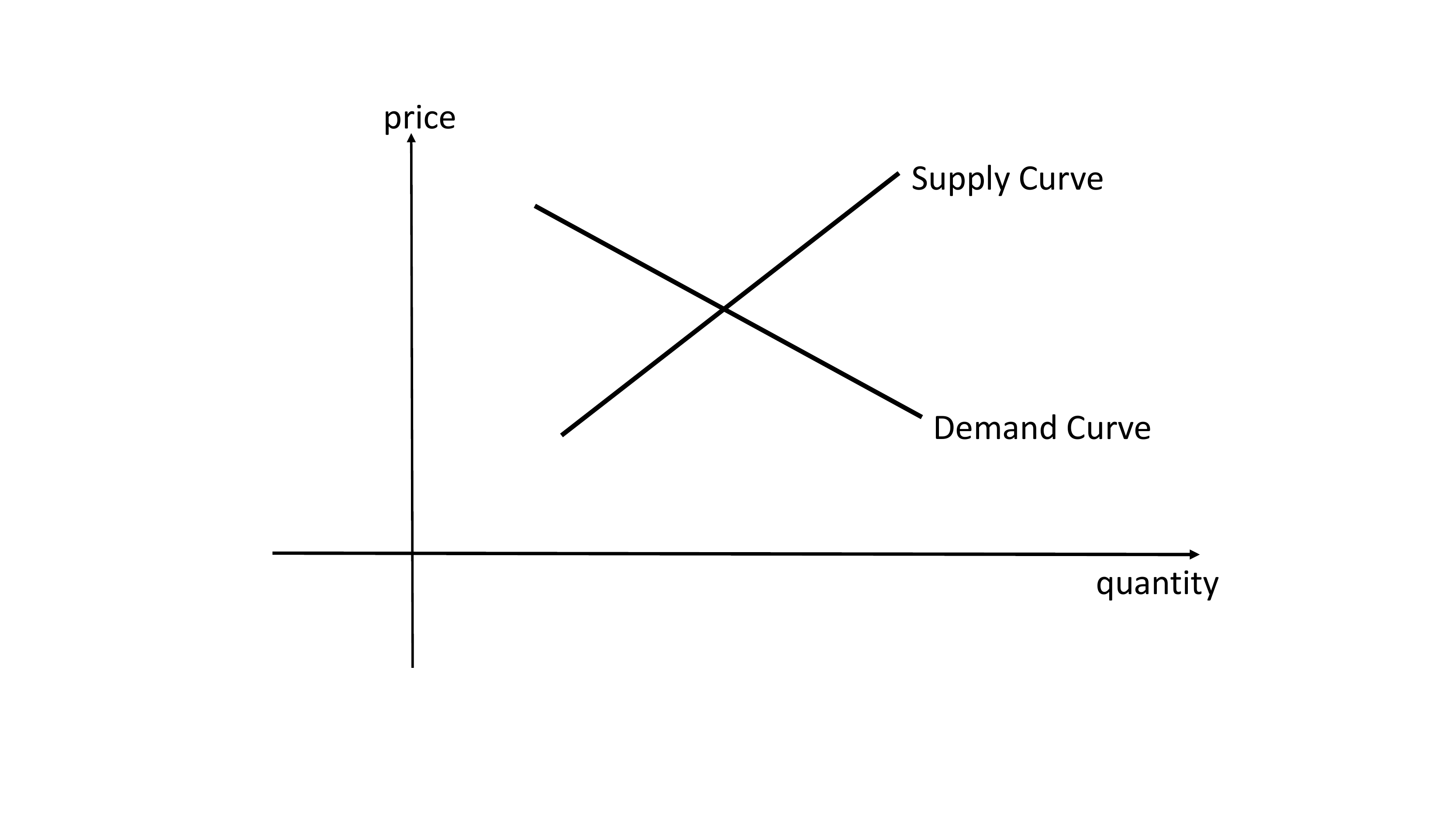}
\caption{The demand curve and supply curve intersecting at the market equilibrium.}
\label{dse}
\end{figure}

\begin{figure}[ht]
\centering
\includegraphics[width=4.5in,height=3in]{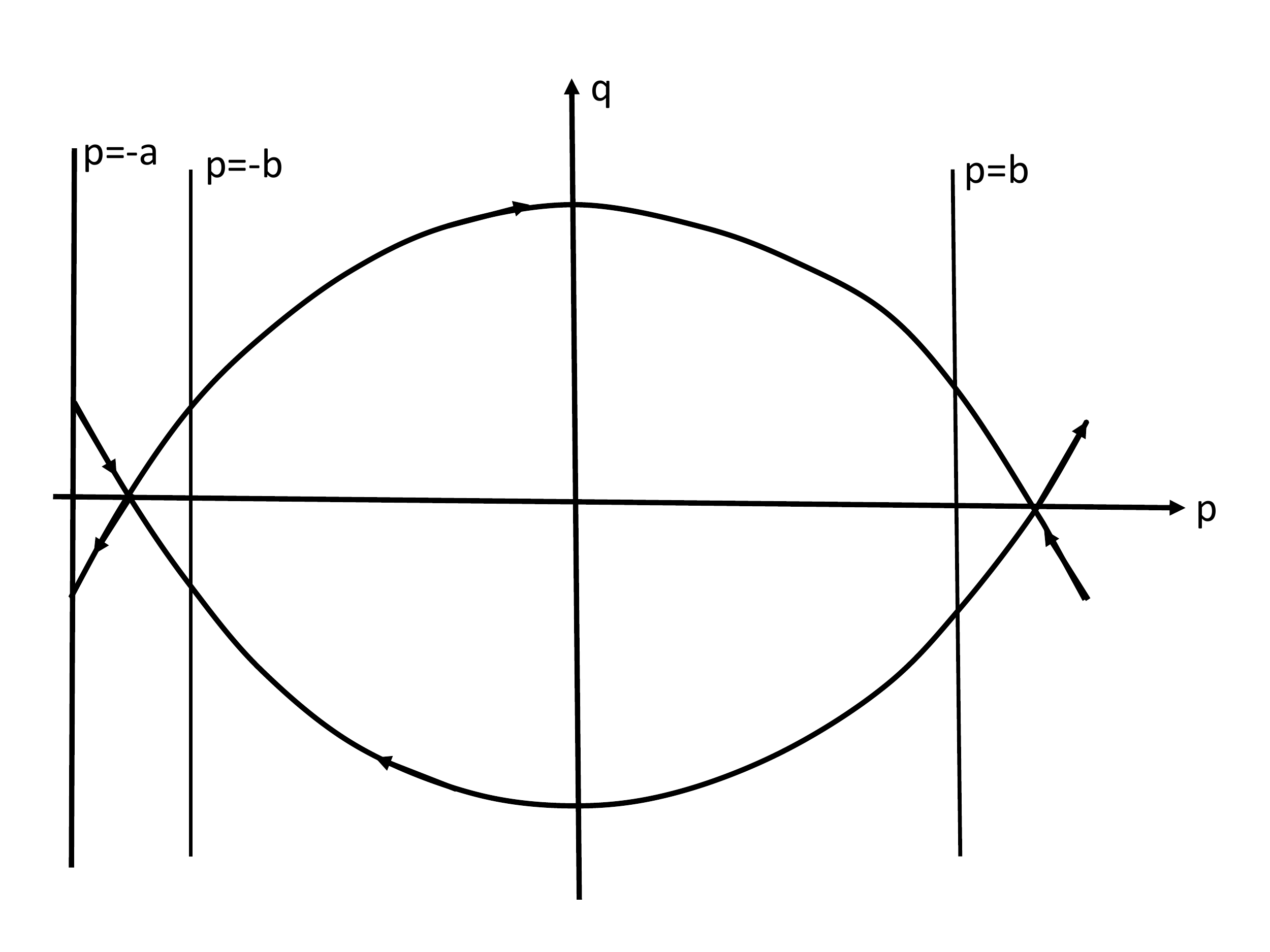}
\caption{The heteroclinic cycle. $a=P_d$ and the dynamics cannot go beyond $p=-a$ to the left since the price $P$ cannot be negative. $b=\sqrt{\frac{1}{\be_1}}$ and $p=\pm b$ are the lines across which the changes of demand switch signs.} 
\label{phd}
\end{figure}

\begin{figure}[ht]
\centering
\includegraphics[width=4.5in,height=3in]{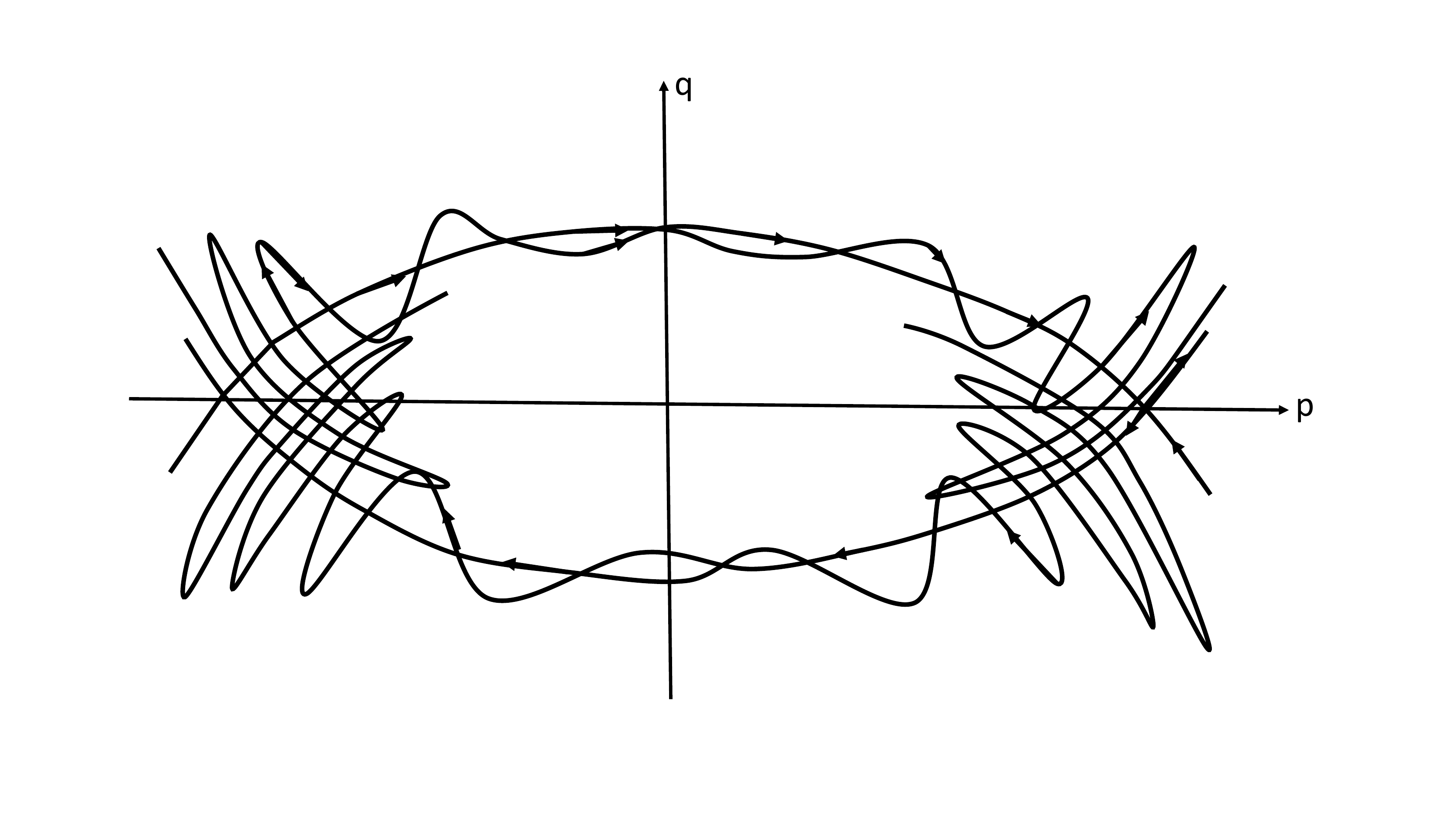}
\caption{The transversal intersection between the broken heteroclinic orbits, forming a Poincar\'e net.}
\label{tii}
\end{figure}

\begin{figure}[ht] 
\centering
\subfigure[$\dl = 0.01, a=0.25$]{\includegraphics[width=2.3in,height=2.3in]{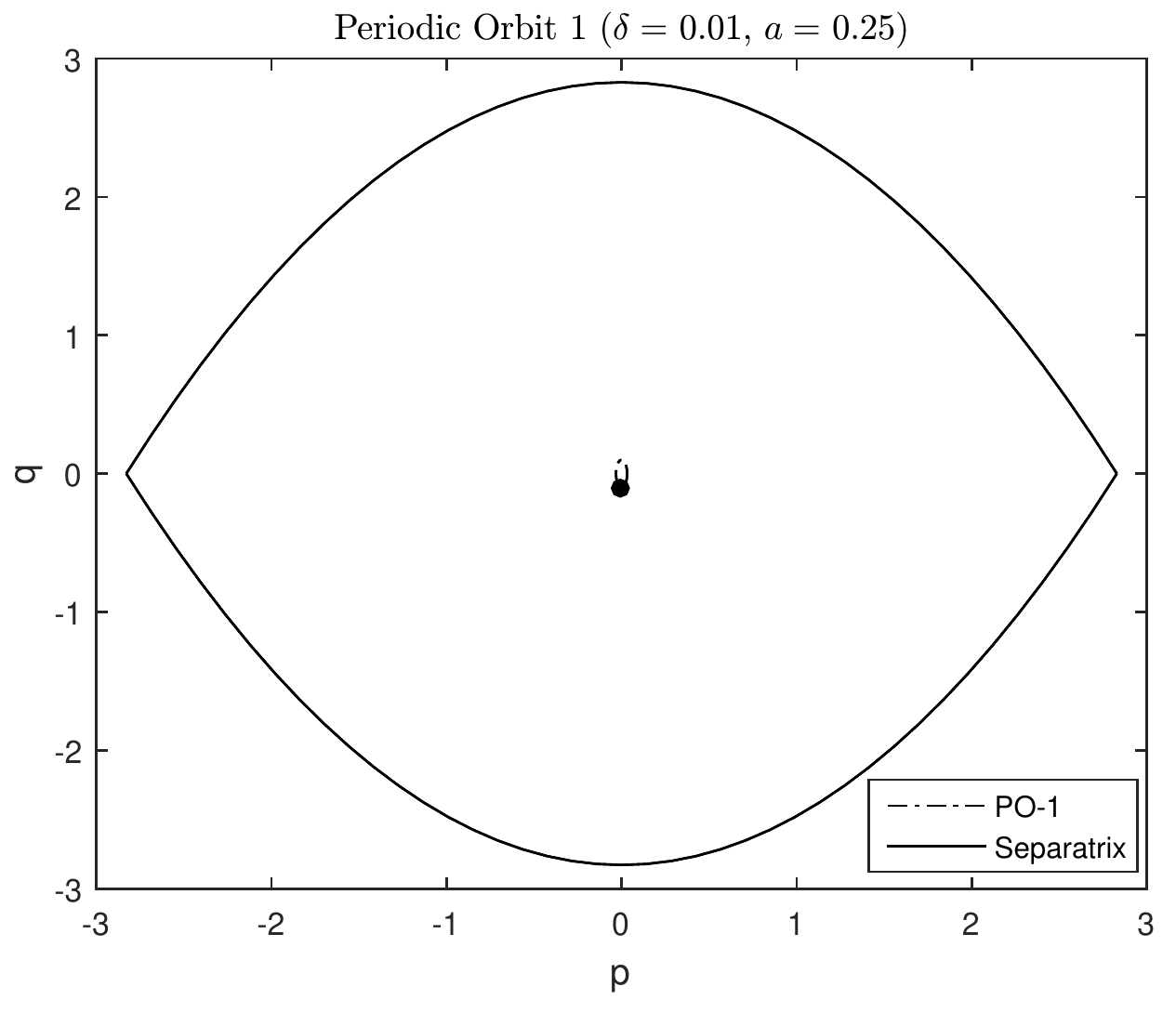}}
\subfigure[$\dl = 0.01, a=0.35$]{\includegraphics[width=2.3in,height=2.3in]{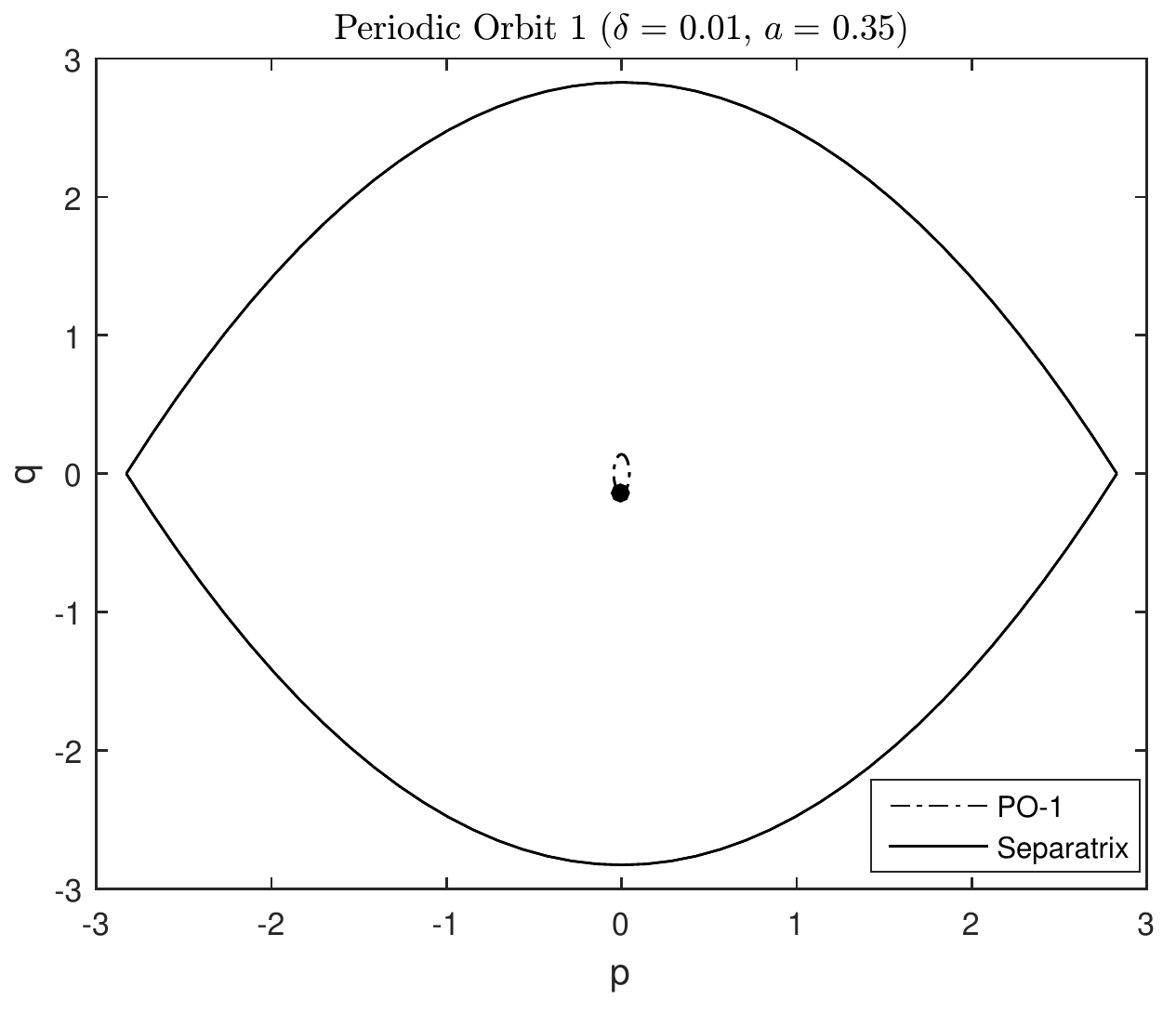}}
\subfigure[$\dl = 0.01, a=0.25$]{\includegraphics[width=2.3in,height=2.3in]{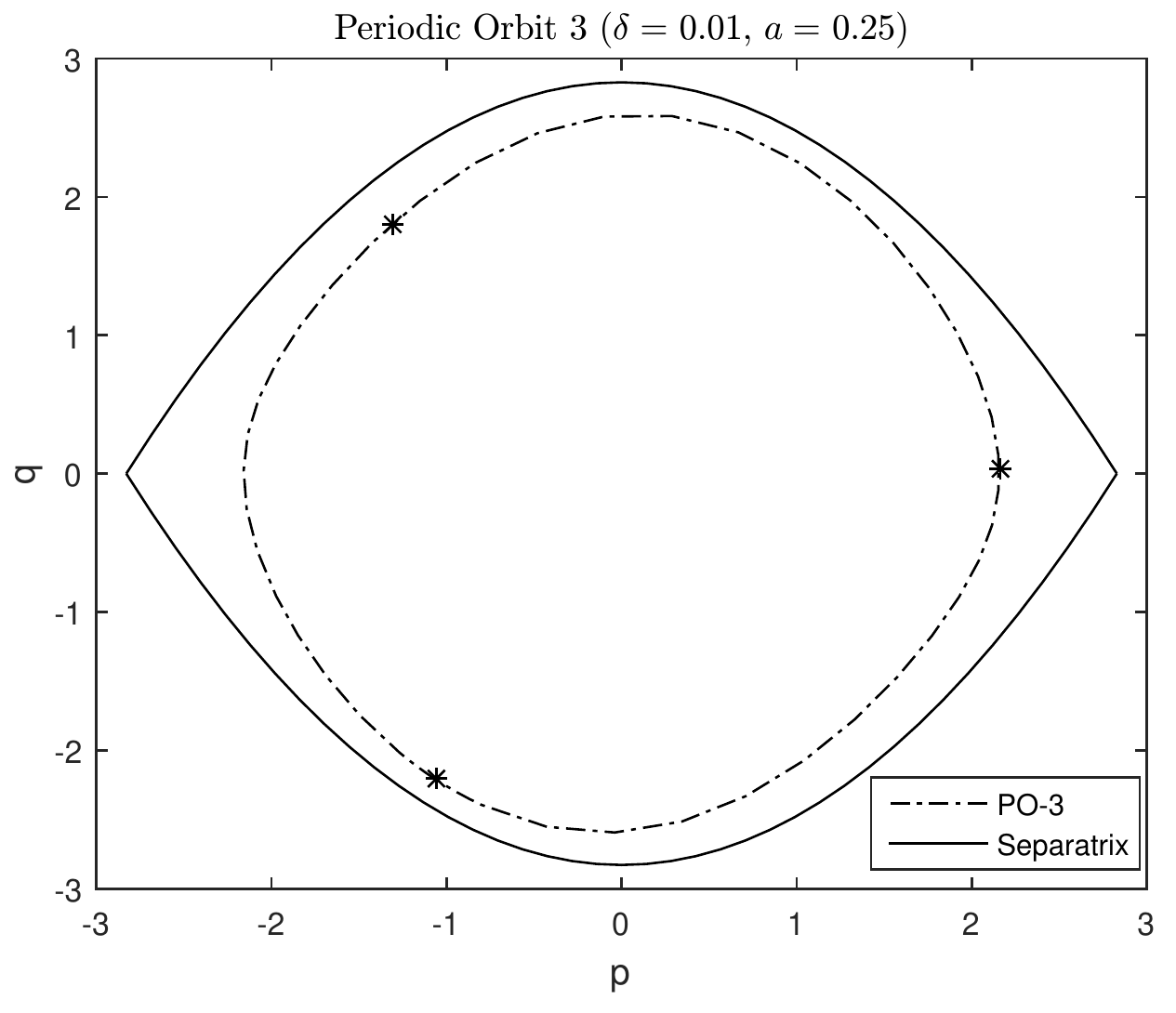}}
\subfigure[$\dl = 0.01, a=0.35$]{\includegraphics[width=2.3in,height=2.3in]{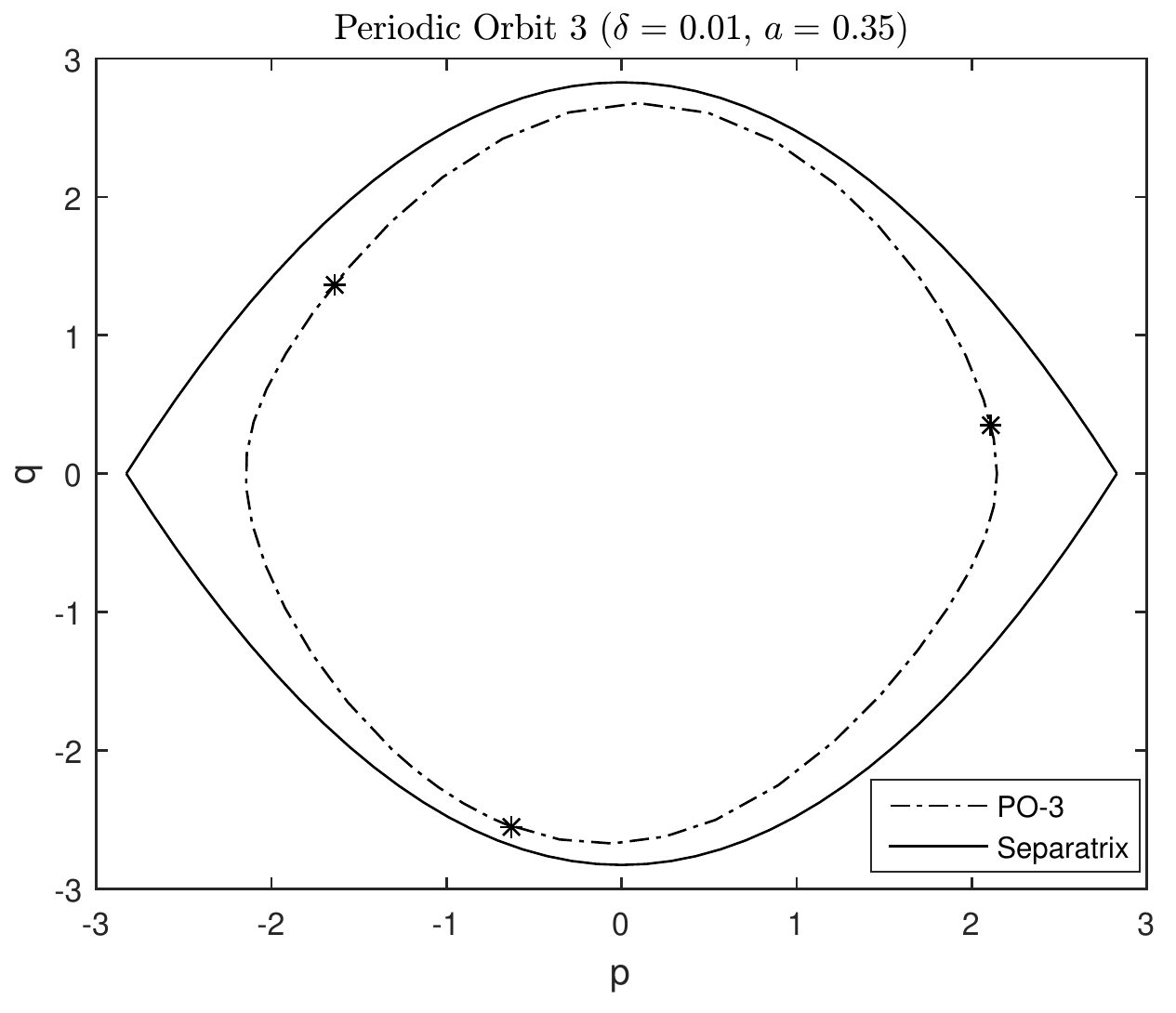}}
\caption{Periodic orbit attractors for values of $a$ below and above its critical value $0.2656$ 
predicted by Melnikov integral, the dot and stars are images under the Poincar\'e period map. The separatrix frame is the one in Fig.\ref{phd}.}
\label{POA1}
\end{figure}

\begin{figure}[ht] 
\centering
\subfigure[$\dl = 0.1, a=2.6$]{\includegraphics[width=2.3in,height=2.3in]{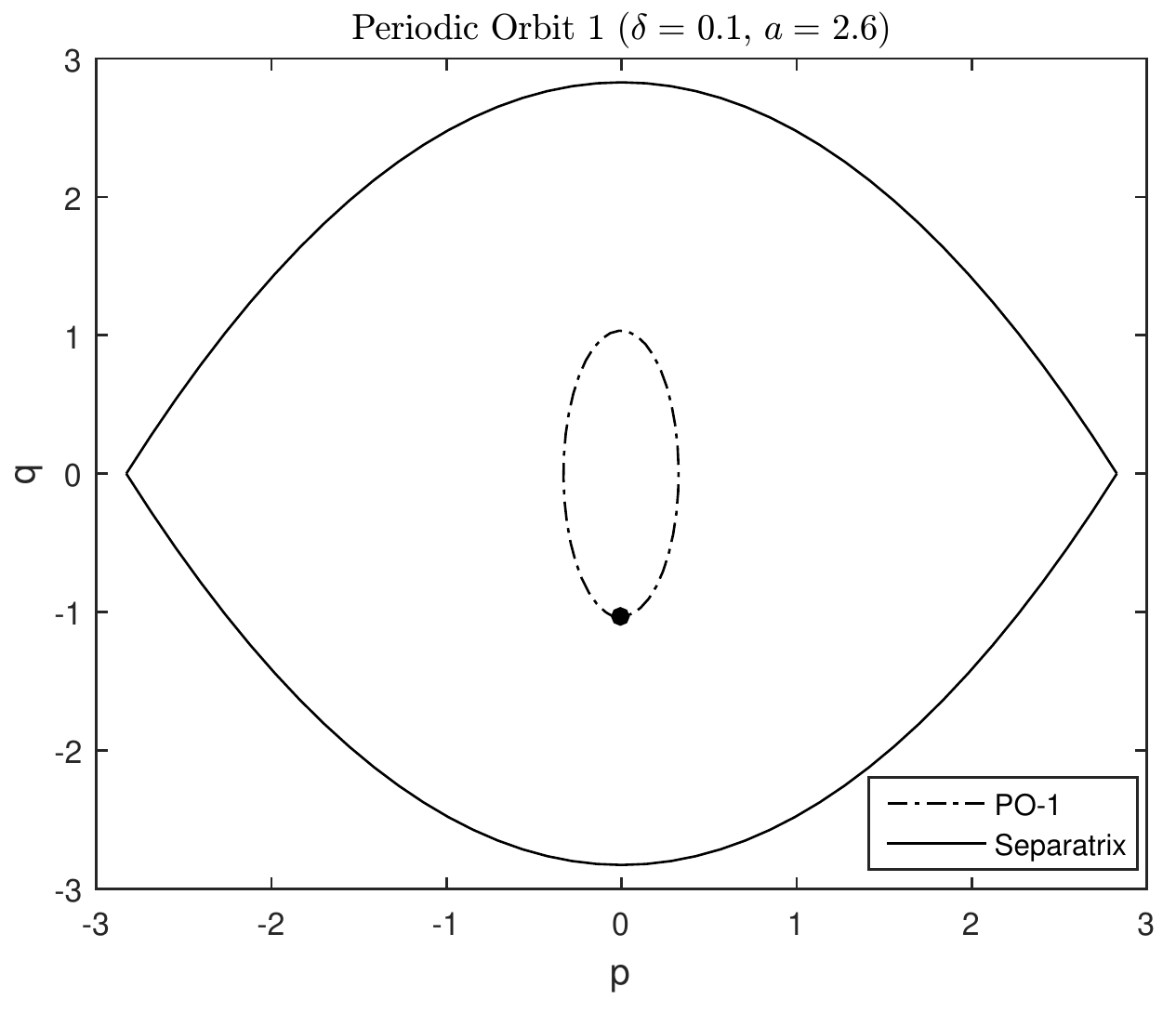}}
\subfigure[$\dl = 0.1, a=3.5$]{\includegraphics[width=2.3in,height=2.3in]{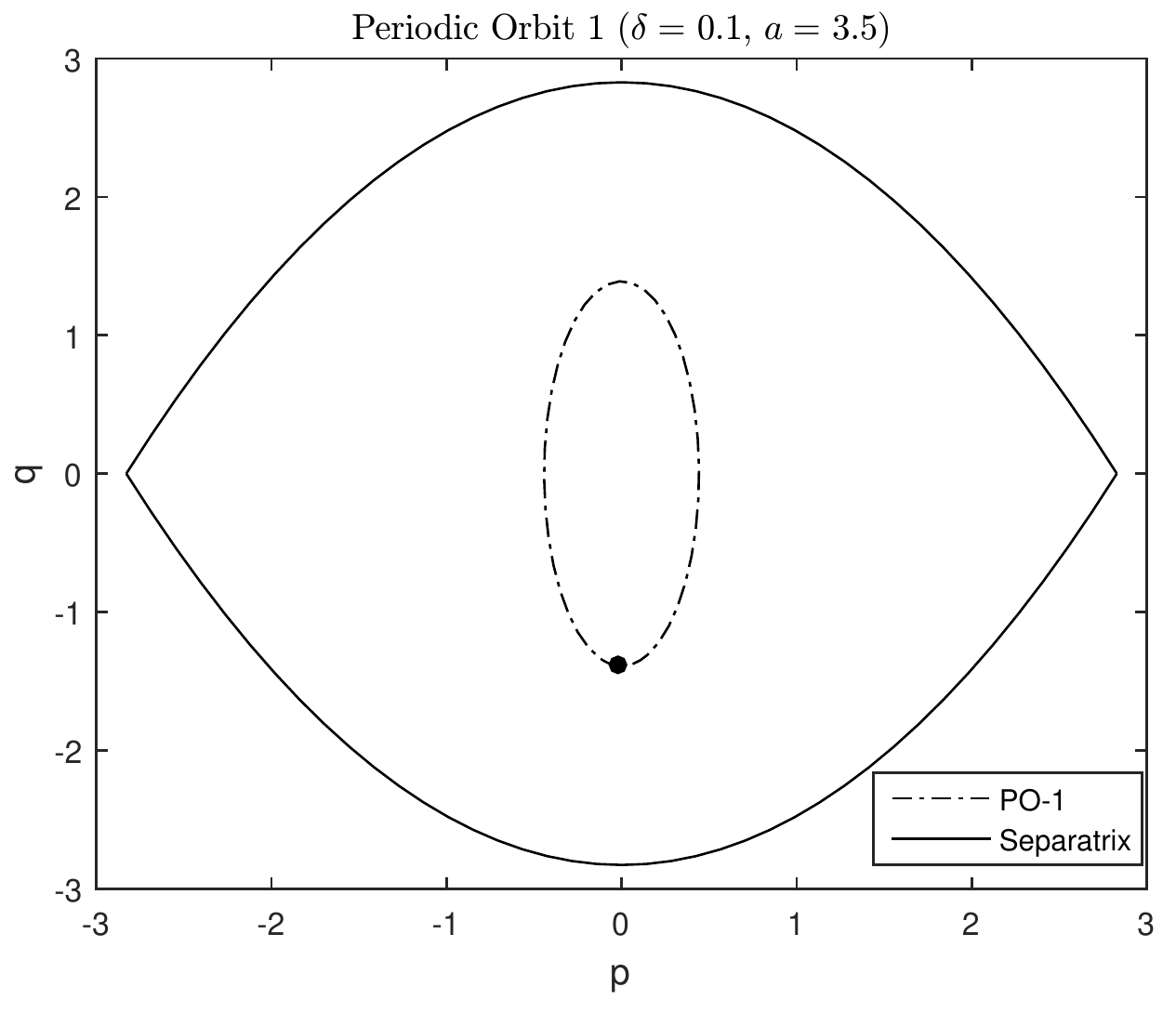}}
\subfigure[$\dl = 0.1, a=2.6$]{\includegraphics[width=2.3in,height=2.3in]{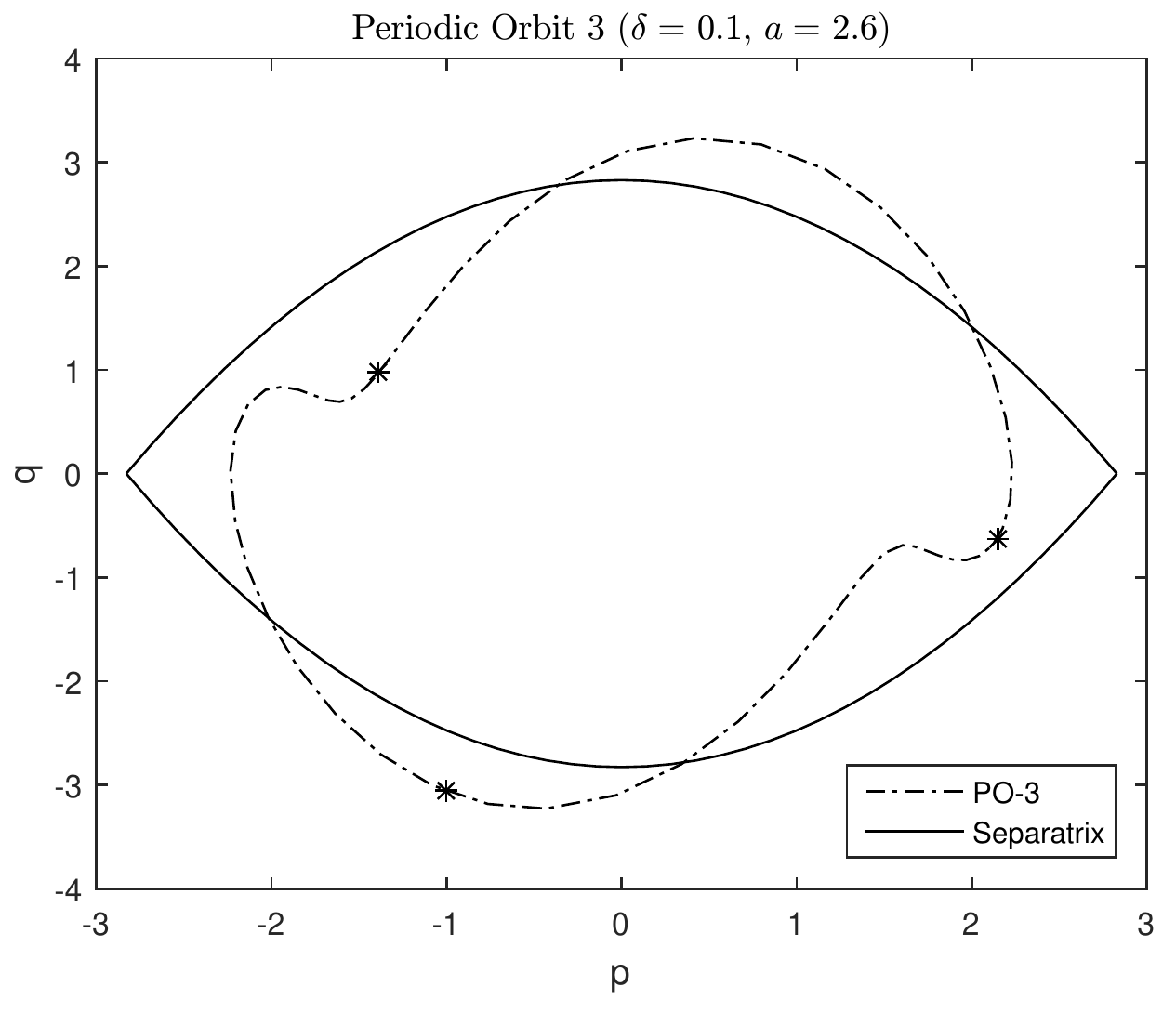}}
\subfigure[$\dl = 0.1, a=3.5$]{\includegraphics[width=2.3in,height=2.3in]{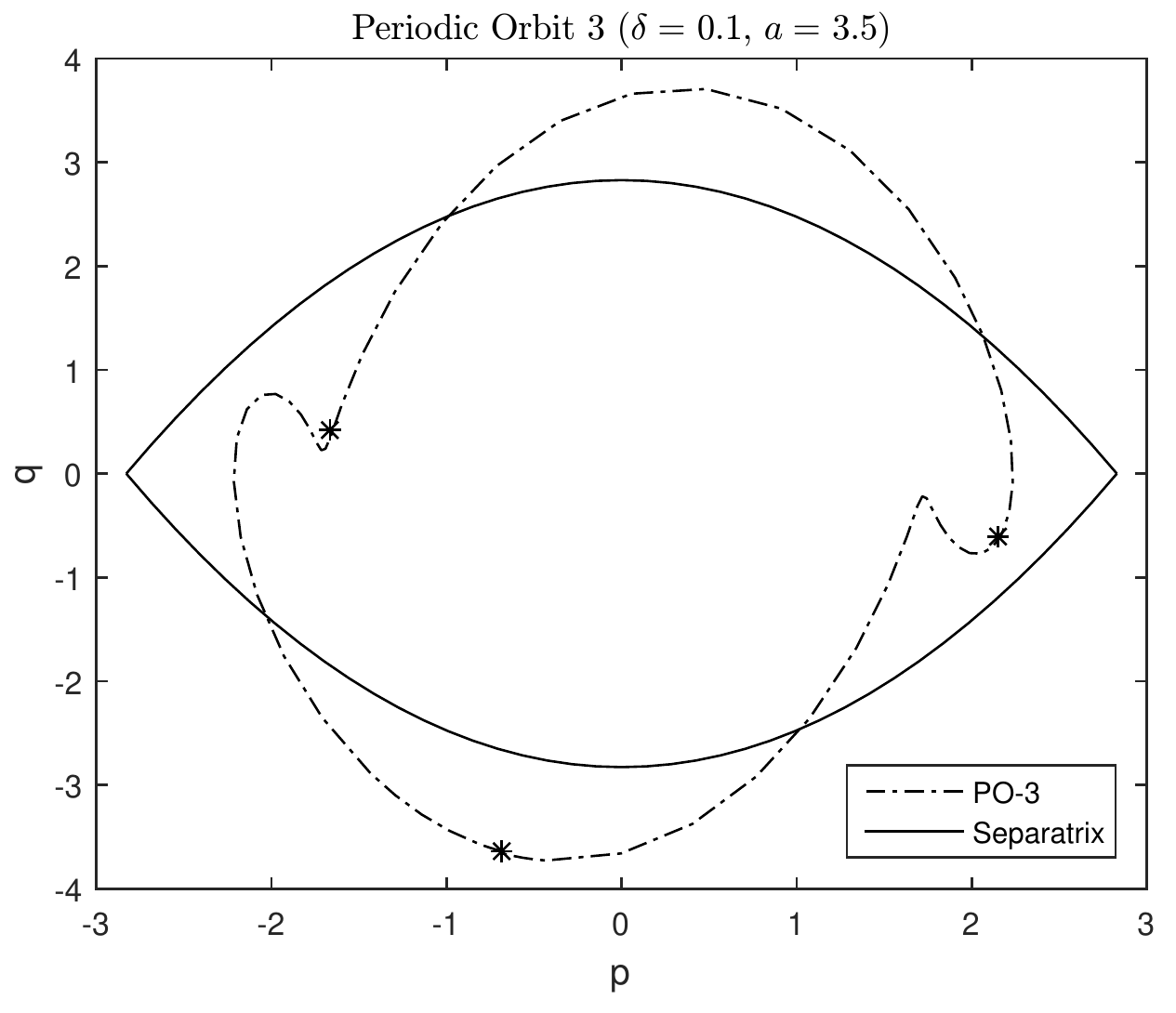}}
\caption{Periodic orbit attractors for values of $a$ below and above its critical value $2.656$ 
predicted by Melnikov integral, the dot and stars are images under the Poincar\'e period map. The separatrix frame is the one in Fig.\ref{phd}.}
\label{POA2}
\end{figure}

\begin{figure}[ht] 
\centering
\subfigure[$\dl = 0.1, a=2.6$]{\includegraphics[width=2.3in,height=2.3in]{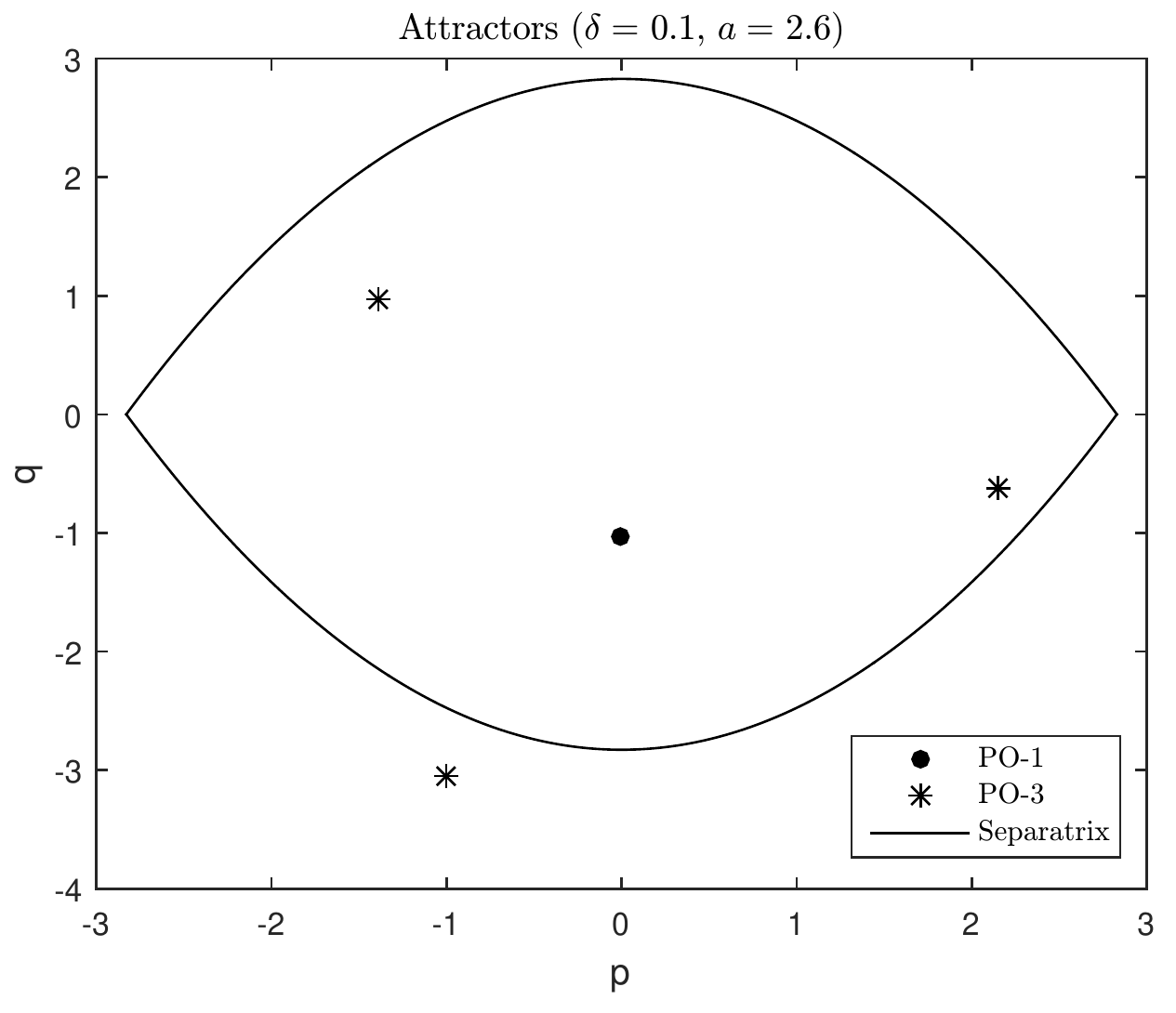}}
\subfigure[$\dl = 0.1, a=3.5$]{\includegraphics[width=2.3in,height=2.3in]{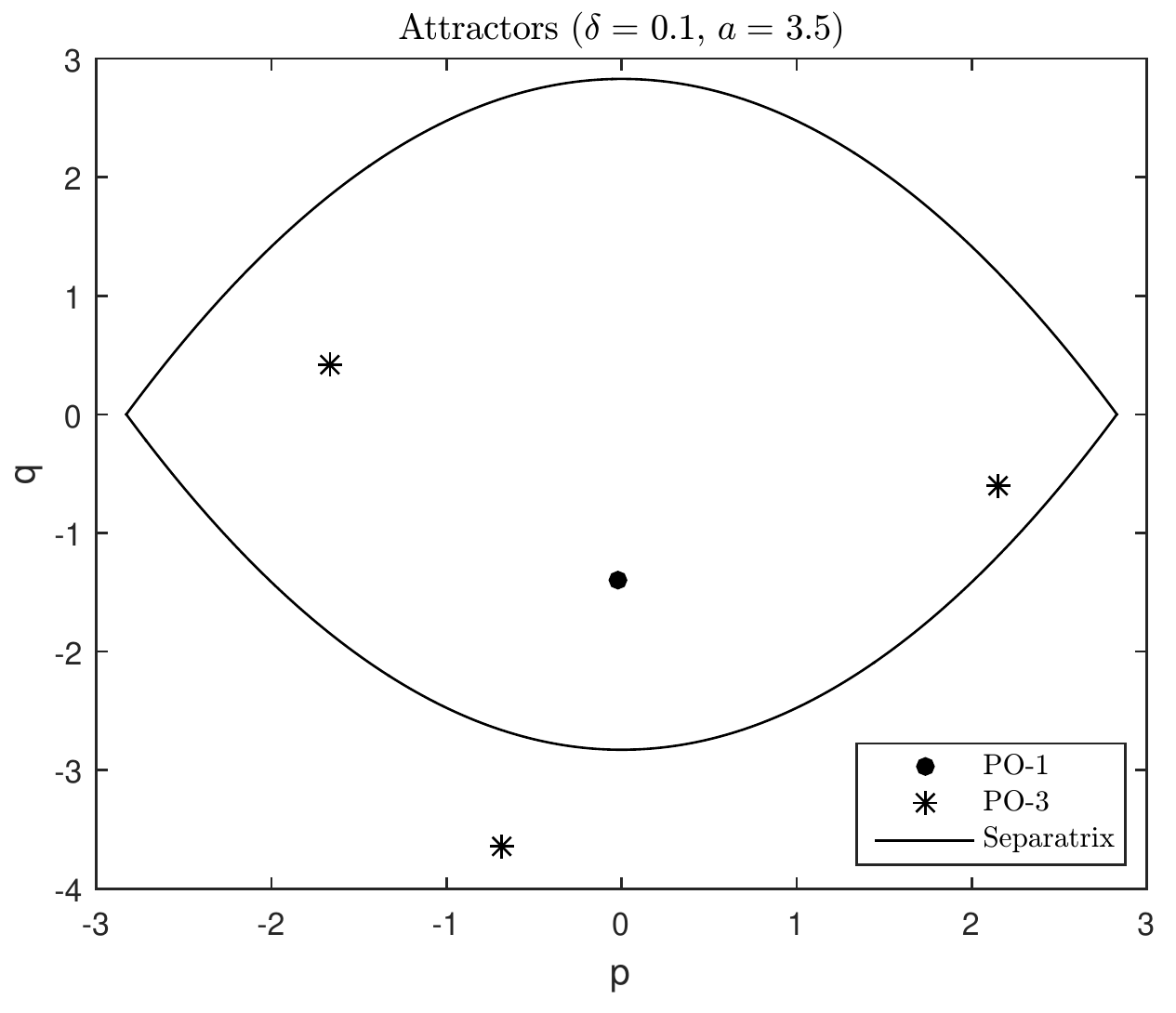}}
\subfigure[$\dl = 0.1, a=2.6$]{\includegraphics[width=2.3in,height=2.3in]{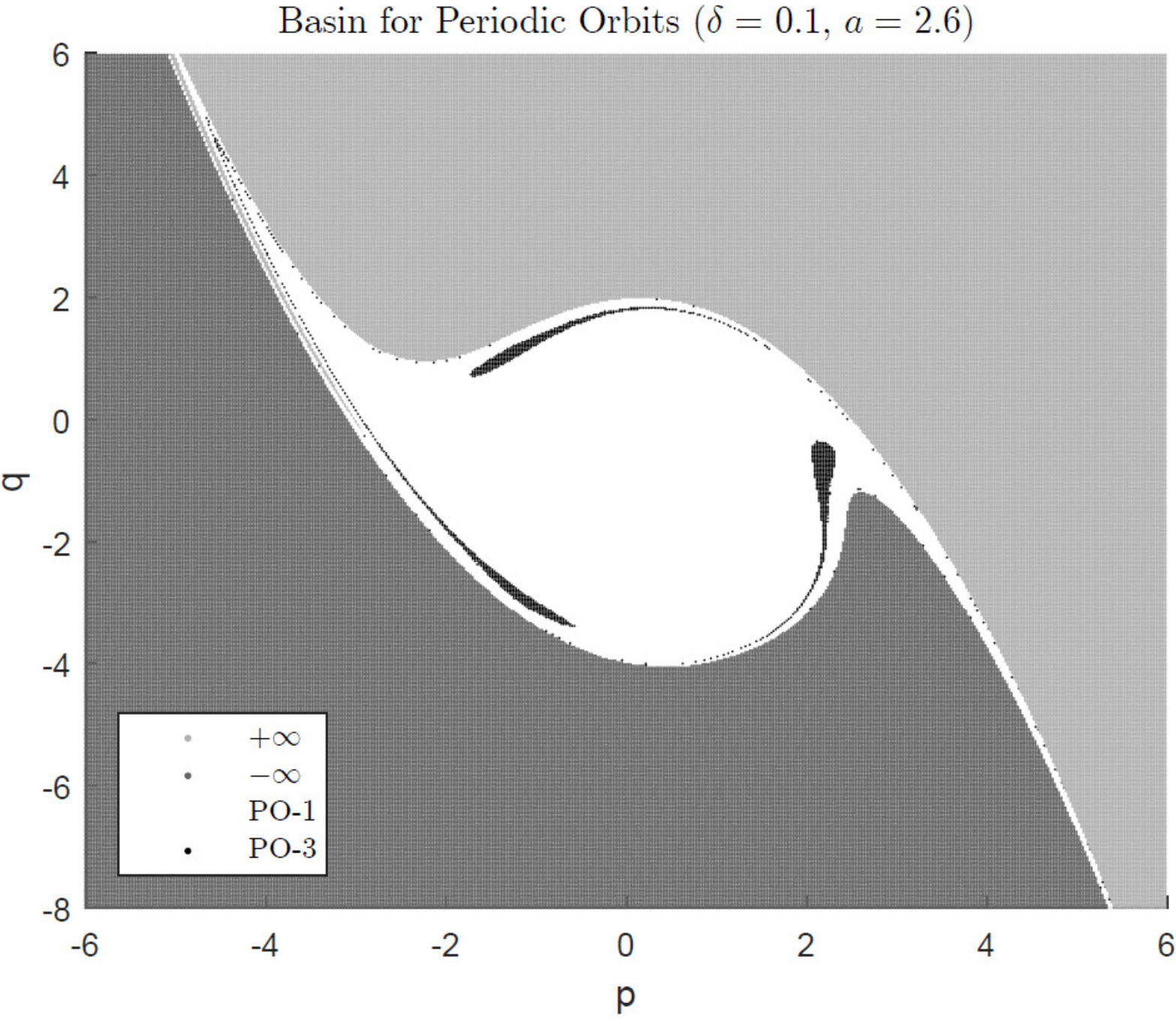}}
\subfigure[$\dl = 0.1, a=3.5$]{\includegraphics[width=2.3in,height=2.3in]{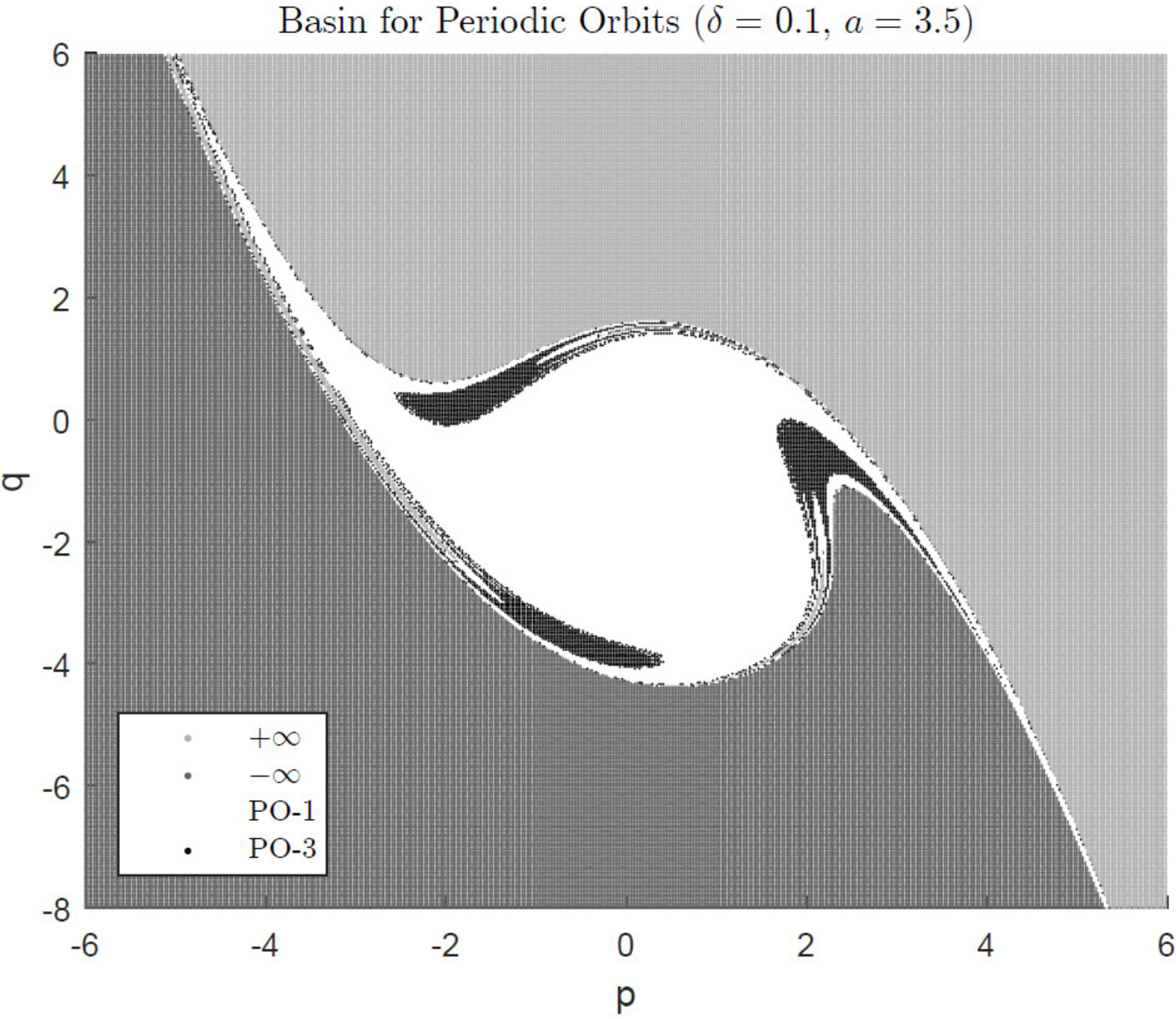}}
\caption{Periodic attractors under the Poincar\'e period map for values of $a$ below and above its critical value $2.656$ predicted by Melnikov integral, the dot represents the period 1 attractor and the 3 stars represents the period 3 attractor. The separatrix frame is the one in Fig.\ref{phd}. In the basin 
of attraction figures, the three leaves form the basin of attraction for the period 3 attractor, the 
central white region is the basin of attraction for the period 1 attractor, the upper region is the basin of attraction for positive infinity ($p,q$) = ($+\infty, +\infty$), and the lower region is the basin of attraction for negative infinity ($p,q$) = ($-\infty, -\infty$).}
\label{POA3}
\end{figure}

\begin{figure}[ht] 
\centering
\subfigure[$\dl = 0.1, a=2.4$]{\includegraphics[width=2.3in,height=2.3in]{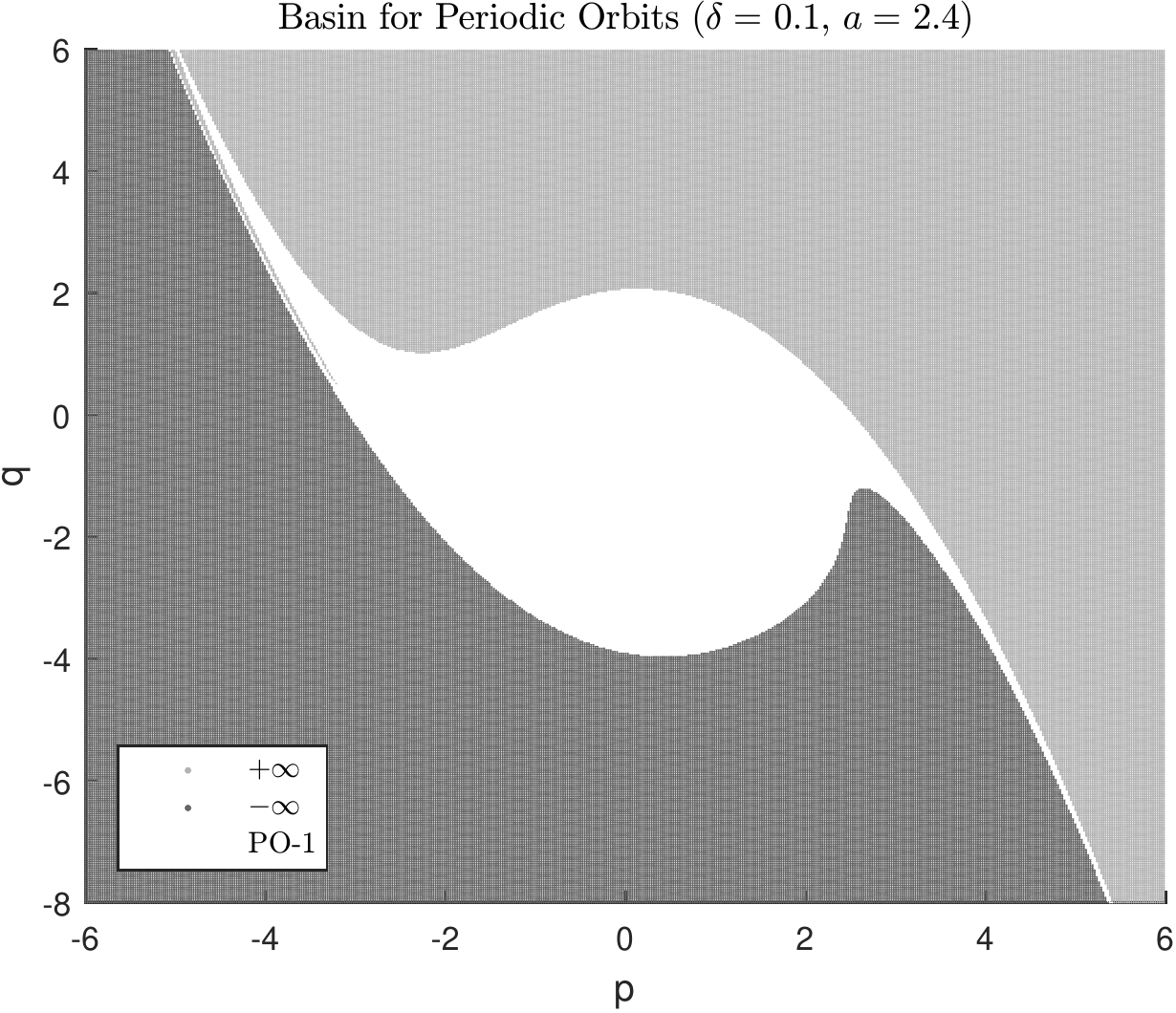}}
\subfigure[$\dl = 0.1, a=5$]{\includegraphics[width=2.3in,height=2.3in]{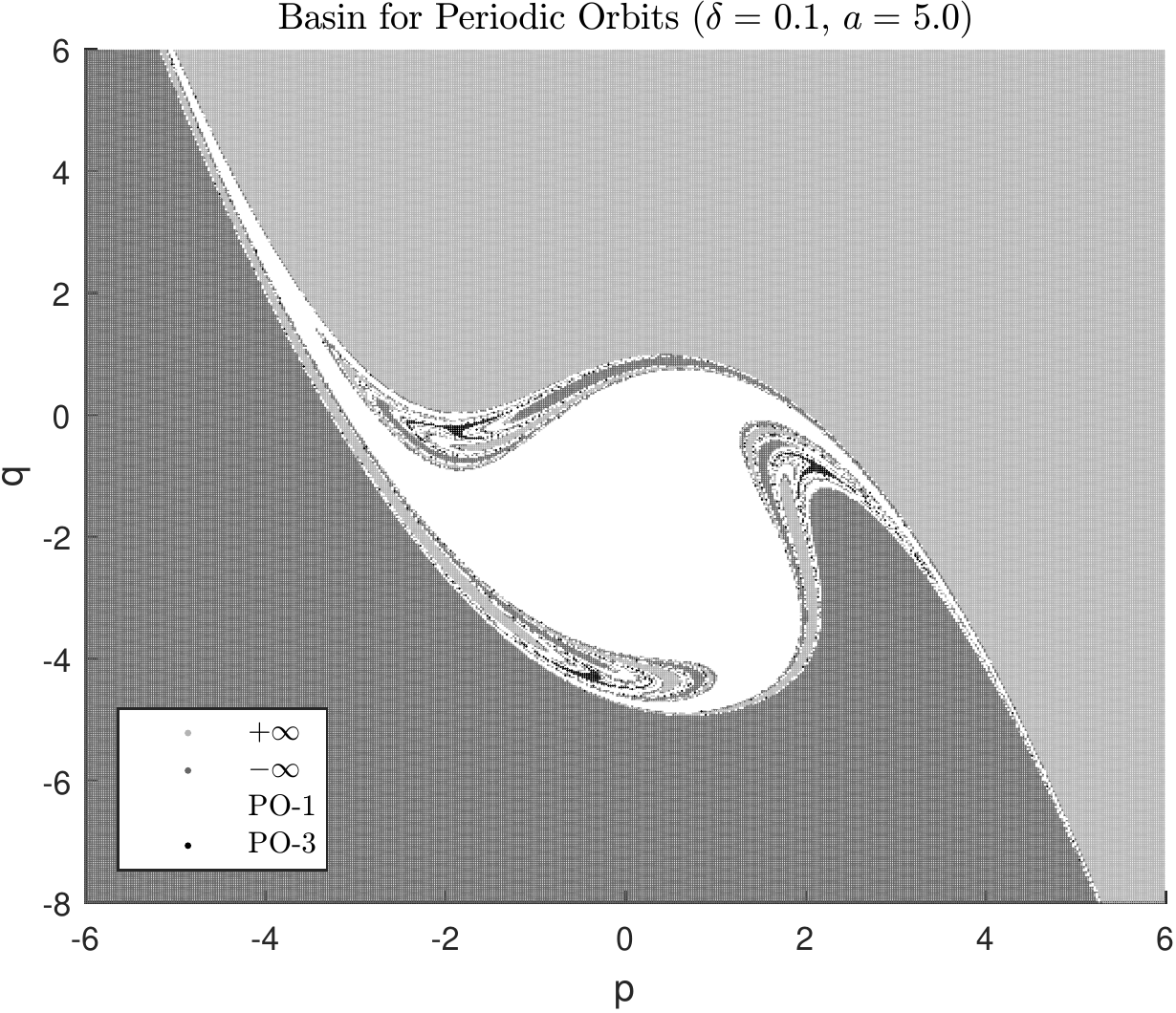}}
\subfigure[$\dl = 0.1, a=5$ zoomed]{\includegraphics[width=2.3in,height=2.3in]{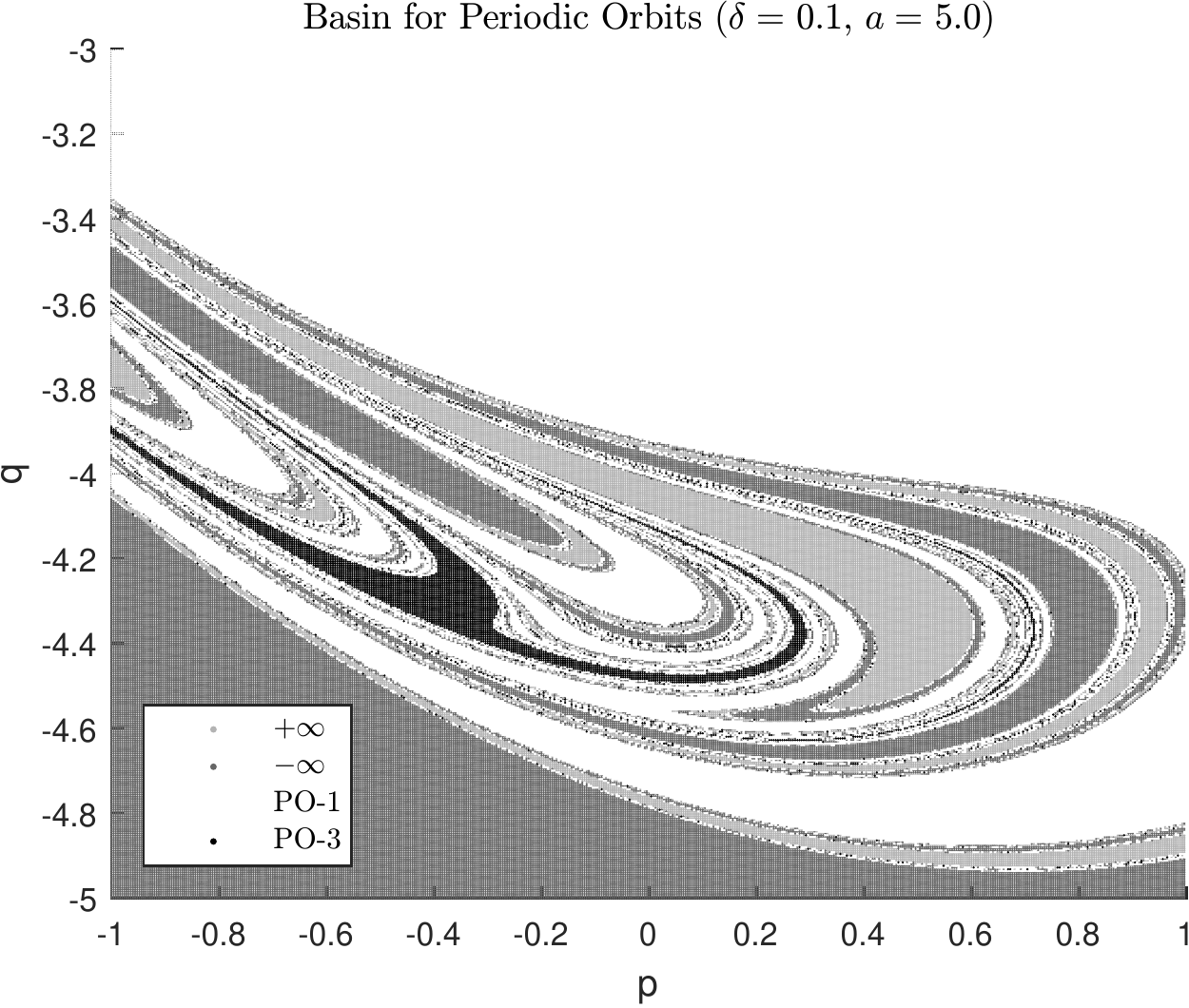}}
\caption{(a). The basins of attraction without period 3 attractor. (b). Fractal basin boundaries among all 
attractors. (c). Zoomed in picture of the lower left leaf in (b).}
\label{FBB}
\end{figure}

\begin{figure}[ht] 
\centering
\subfigure[$\dl = 0.1, a=6.4$]{\includegraphics[width=2.3in,height=2.3in]{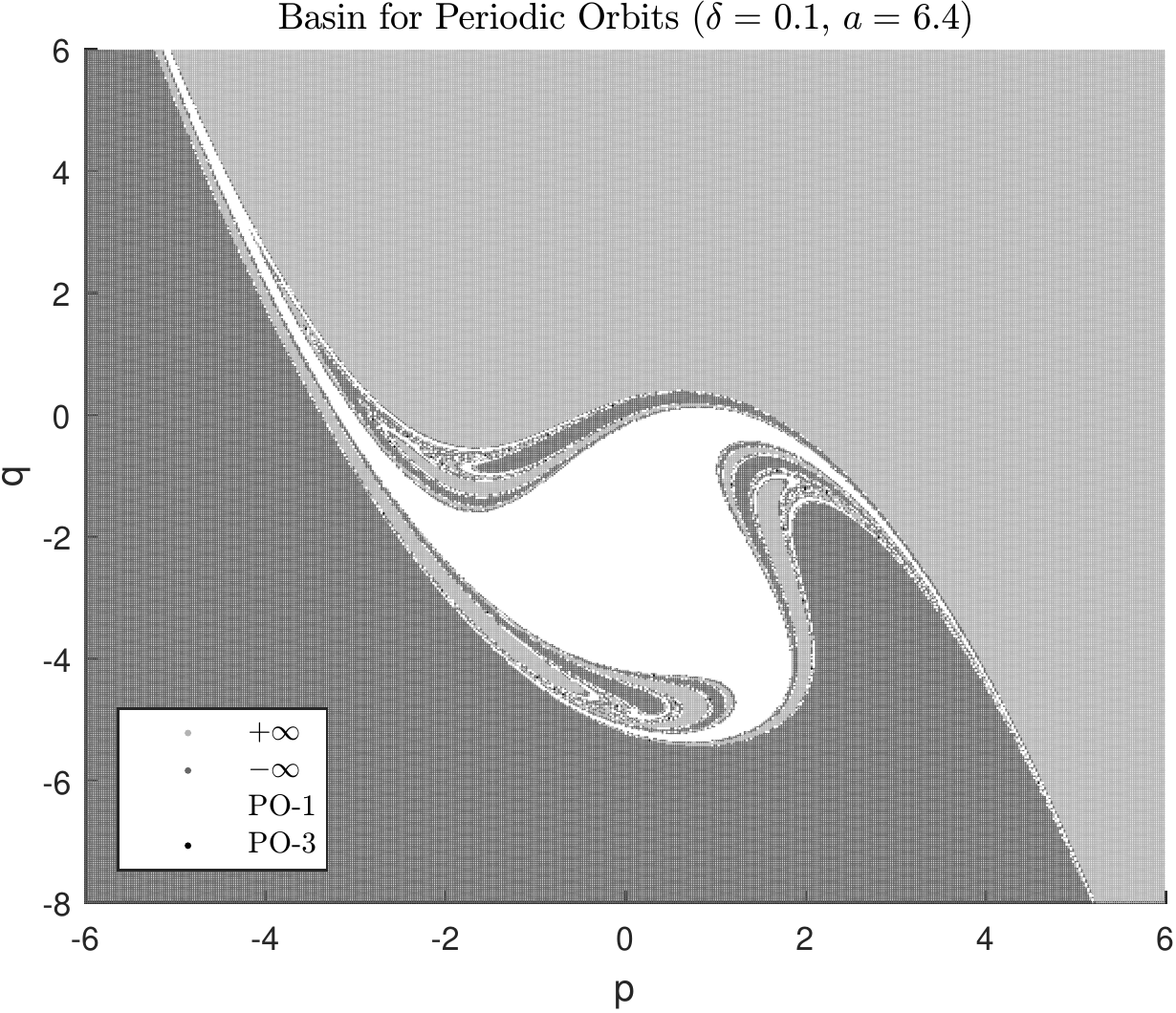}}
\subfigure[$\dl = 0.1, a=6.4$ zoomed]{\includegraphics[width=2.3in,height=2.3in]{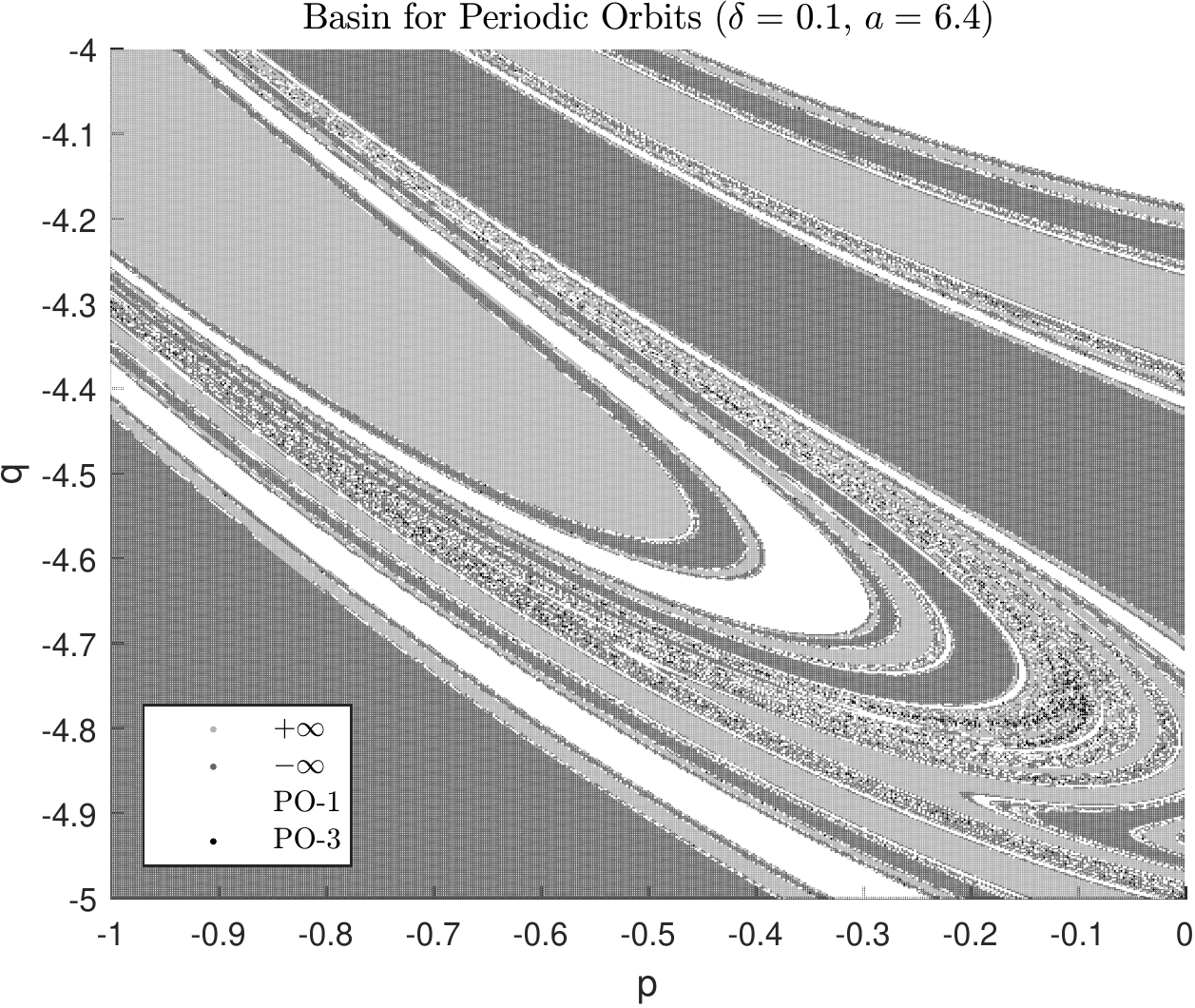}}
\subfigure[$\dl = 0.1, a=6.5$]{\includegraphics[width=2.3in,height=2.3in]{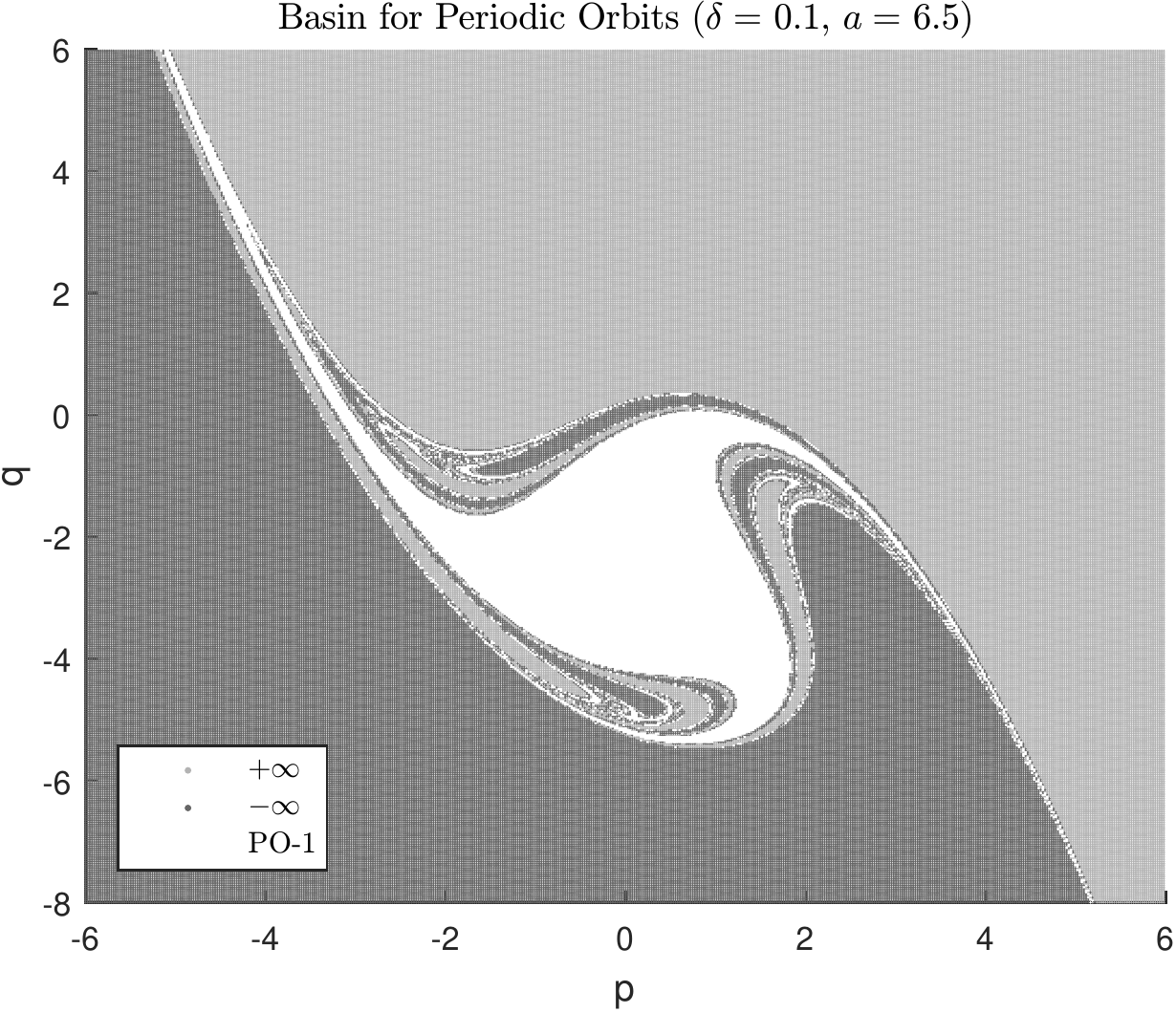}}
\subfigure[$\dl = 0.1, a=6.5$ zoomed]{\includegraphics[width=2.3in,height=2.3in]{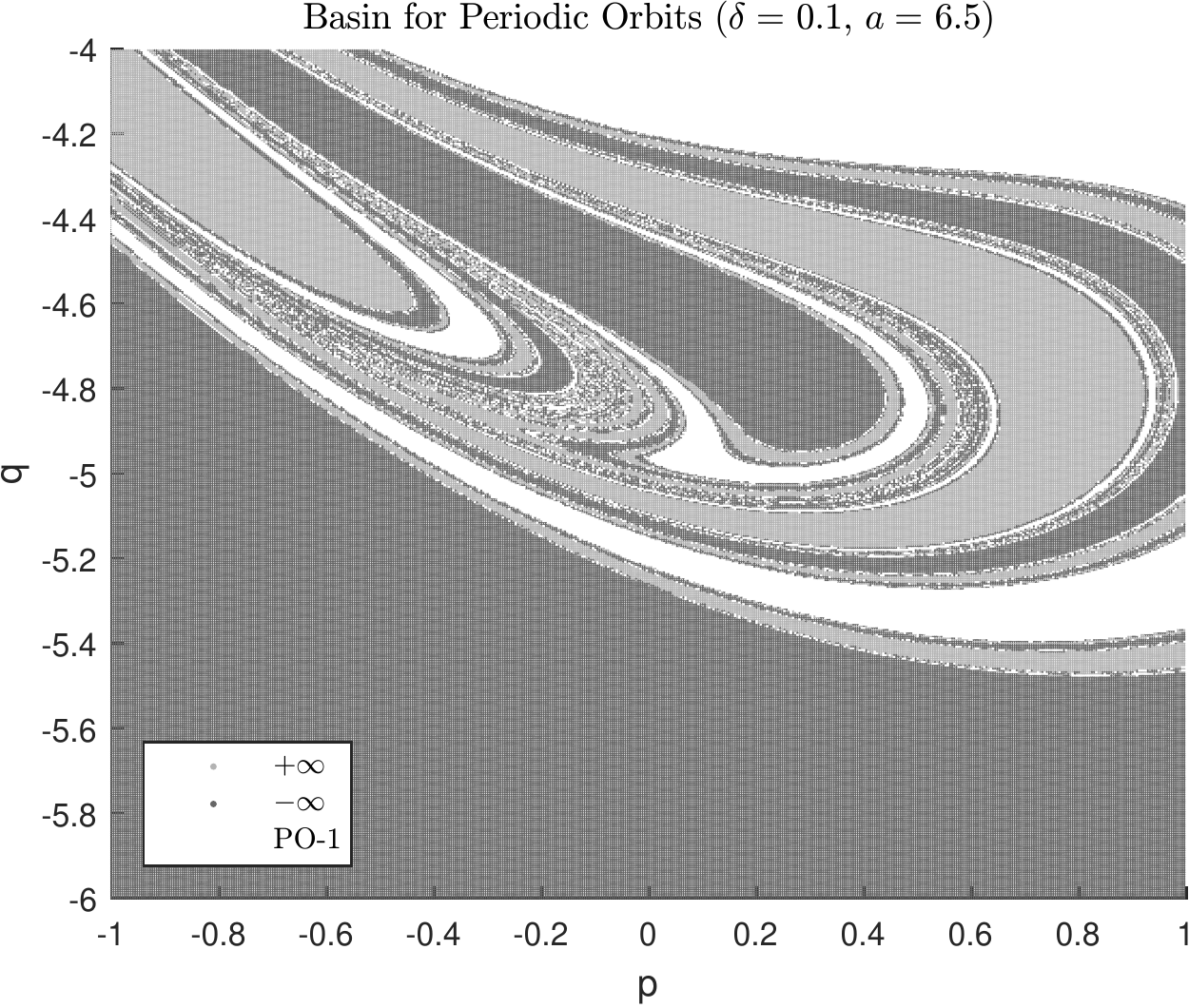}}
\caption{(a). Right before the basin of attraction of period 3 attractor disappears. (b). Zoomed in picture of the lower left leaf in (a). (c). The basin of attraction of period 3 attractor totally disappears, still having 
fractal basin boundaries. (d). Zoomed in picture of the lower left leaf in (c).}
\label{FCE}
\end{figure}

\section{Analysis of the simple specific model}

\subsection{Integrable dynamics}

When $\dl = a = 0$, the system (\ref{am1})-(\ref{am2}) is an integrable Hamiltonian system
\begin{eqnarray}
&& \frac{dp}{dt} = \frac{\pa H}{\pa q} , \label{ham1} \\
&& \frac{dq}{dt} =- \frac{\pa H}{\pa p} , \label{ham2} 
\end{eqnarray}
where
\[
H = \frac{1}{2} \al q^2 +\frac{1}{2} (\be +\ga ) p^2 - \frac{1}{4}\be \be_1 p^4 .
\]
There are three fixed points in the system (\ref{ham1})-(\ref{ham2}) when
\begin{equation}
\frac{\be +\ga }{\be } < \be_1 P_d^2 \label{sc2}
\end{equation}
which is stronger than (\ref{sc}). The three fixed points are 
\begin{equation}
(0,0), \quad \left ( \pm \sqrt{\frac{\be +\ga}{\be \be_1}} , 0 \right ) . \label{fpts}
\end{equation}
The first fixed point represents the market equilibrium, the negative $p$-value fixed point is named the saturation fixed point, and the positive $p$-value fixed point is 
named the collectability fixed point. The market equilibrium is a neutrally stable center. The eigenvalues of the market equilibrium under the linearized dynamics of (\ref{ham1})-{\ref{ham2}) are 
\[
\la = \pm i \sqrt{\al (\be +\ga )}.
\]
The saturation and collectability fixed points are unstable saddles with eigenvalues
\[
\la = \pm \sqrt{2\al (\be +\ga )} .
\]
The phase plane diagram of (\ref{ham1})-(\ref{ham2}) is shown in Figure \ref{phd}. The minimal value of $p$ is $-P_d$ (since the price $P$ can not be negative). As $p$ decreases across 
$-\sqrt{\frac{1}{\be_1}}$, demand switches from increasing to decreasing (cf: (\ref{sm})), and the dynamics of (\ref{ham1})-(\ref{ham2}) reaches a (unstable) equilibrium at 
$p = -\sqrt{\frac{\be +\ga }{\be \be_1}}$ and $D=S$. As $p$ increases  across $\sqrt{\frac{1}{\be_1}}$, demand switches from decreasing to increasing (cf: (\ref{sm})), and the dynamics 
of (\ref{ham1})-(\ref{ham2}) reaches a (unstable) equilibrium at $p = \sqrt{\frac{\be +\ga }{\be \be_1}}$ and $D=S$. In terms of the originial variables, at the market equilibrium (\ref{fpts}),
\[
P = P_d, \quad D=S= \text{ constant}. 
\]
At the saturation fixed point (\ref{fpts}),
\[
P=P_d - \sqrt{\frac{\be +\ga }{\be \be_1}},\quad  D=S,\quad  \frac{dD}{dt} = \frac{dS}{dt} = \text{ negative constant},
\]
thus, the amounts of demand and supply are equal and decrease at the same rate in time. Such a state is unstable. At the collectability fixed point (\ref{fpts}),
\[
P=P_d + \sqrt{\frac{\be +\ga }{\be \be_1}}, \quad D=S, \quad \frac{dD}{dt} = \frac{dS}{dt} = \text{ positive constant},
\]
thus, the amounts of demand and supply are equal and increase at the same rate in time. Such a state is unstable. 

On the phase plane (Figure \ref{phd}), connecting the saturation and collectability fixed points is a heteroclinic cycle of two heteroclinic orbits. The upper heteroclinic orbit has the expression
\begin{eqnarray}
&& p = A \tanh (\Om t+t_0), \label{heto1} \\
&& q= \frac{A \Om }{\al } \text{ sech}^2 (\Om t +t_0 ), \label{heto2} 
\end{eqnarray}
where $t_0$ is a parameter,
\[
A = \sqrt{\frac{\be +\ga }{\be \be_1}}, \quad \Om = \sqrt{\frac{1}{2} \al (\be +\ga )} .
\]
The lower heteroclinic orbit is given by ($-p,-q$) where ($p,q$) is the upper heteroclinic orbit (\ref{heto1})-(\ref{heto2}). 

\subsection{Chaotic dynamics}

When $\dl \neq 0$ and $a \neq 0$, the dynamics of (\ref{am1})-(\ref{am2}) is not integrable, and we will show that it is chaotic via a Melnikov integral and Shadowing Lemma. The Melnikov integral is given by \cite{Li04},
\[
M = \int_{-\infty}^{+\infty} dH
\]
evaluated along the heteroclinic orbit (\ref{heto1})-(\ref{heto2}). Thus
\begin{eqnarray*}
M &=& \al \int_{-\infty}^{+\infty} [-\dl q^2 + a q \sin (\om_1 t) ] dt \\
&=& -\frac{\dl A^2 \Om }{\al }  \int_{-\infty}^{+\infty} \text{ sech}^4 \tau d \tau + aA  \int_{-\infty}^{+\infty} \text{ sech}^2 \tau \sin \left (\frac{\om_1}{\Om}\tau -\frac{\om_1}{\Om}t_0 \right )d\tau \\
&=& -\frac{4\dl A^2 \Om }{3\al } -2aA \sin \frac{\om_1}{\Om}t_0 \int_{0}^{+\infty} \text{ sech}^2 \tau  \cos \frac{\om_1}{\Om}\tau d\tau . 
\end{eqnarray*}
Setting $M=0$, we get
\begin{equation}
\sin \frac{\om_1}{\Om}t_0 = -\frac{2\dl A \Om }{3\al  a}  \left [ \int_{0}^{+\infty} \text{ sech}^2 \tau  \cos \frac{\om_1}{\Om}\tau d\tau \right ]^{-1} . \label{MR}
\end{equation}
When 
\begin{equation}
|a| >  \frac{2\dl A \Om }{3\al  }  \left [ \int_{0}^{+\infty} \text{ sech}^2 \tau  \cos \frac{\om_1}{\Om}\tau d\tau \right ]^{-1} , \label{TR}
\end{equation}
the Melnikov integral $M$ has infinitely many simple roots given by (\ref{MR}) which imply that the broken pieces of the heteroclinic orbit re-intersect transverally under the Poincar\'e map $F$ of (\ref{am1})-(\ref{am2}) 
(a fact proven mathematically rigorously when $\dl$ and $a$ are small \cite{Li04}) (Fig.\ref{tii}). The intersection points form a transversal heteroclinic cycle under the Poincar\'e map $F$ of (\ref{am1})-(\ref{am2}). 
Then via Shadowing Lemma approach, it is rigorously proved that there is chaos in the dynamics of (\ref{am1})-(\ref{am2}) \cite{Li04}. The next key question is whether or not the chaos is an attractor, and this will be answered by numerical simulations.

\section{Numerical simulation of the simple specific model}

Here we are going to numerically simulate the dynamics of (\ref{am1})-(\ref{am2}) in terms of attractors
and their basins of attraction. We choose the parameters as follows:
\[
\al = 1, \ \be = 1, \ \be_1 = 0.25, \ \ga = 1, \ \om_1 = \pi . 
\]
We leave the other two parameters $\dl$ and $a$ adjustable for different numerical simulations, and recall that 
$\dl = a =0$ corresponds to the integrable dynamics. 
When $\dl =0.01$, the Melnikov integral predicts that when $|a|> 0.2656$, there exists chaos. But this chaos may not be an attractor. Our numerical simulations show that this is indeed the case: the chaos is not an attractor, instead a period-3 attractor (under the Poincar\'e map) exists near the chaos. In fact, there are 
two co-existing periodic attractors (period-1 and period-3 under the Poincar\'e map) as shown in 
Figure \ref{POA1}. In Figure \ref{POA1}, two values of $a$ are chosen $a=0.25$ (below the critical value $0.2656$) and $a=0.35$ (above the critical value $0.2656$). The solid dot is the period-1 attractor under the Poincar\'e map, while the small loop above the dot is the continuous periodic attractor under the dynamics of  
(\ref{am1})-(\ref{am2}). The three stars form the period-3 attractor under the Poincar\'e map, while the loop connecting the three stars is the the continuous periodic attractor under the dynamics of  
(\ref{am1})-(\ref{am2}). For reference, we also plot the separatrix in Figure \ref{phd}. One can see clearly 
that the attractor structure is the same for both cases $a=0.25$ and $a=0.35$. This shows that when the value of $a$ crosses the critical value $a=0.2656$, the structure of the attractors does not change, and chaos is 
not an attractor. The same conclusion holds in Figure \ref{POA2} where $\dl =0.1$ and the critical value is 
$a=2.656$. The period-1 attractor represents the market regular fluctuation near the market equilibrium, while the period-3 attractor represents the market recession (depression) and large growth cycle. Next we will focus on the entire phase space and study the basins of attraction of all the attractors. For $\dl =0.1$, when 
$2.4 < a <6.5$, the period-3 and period-1 attractors coexist, and in this case there are total four attractors,
positive infinity and negative infinity besides the period-3 and period-1 attractors. Positive infinity 
and negative infinity represent market irrational exuberance and flash crash. When $a\leq 2.4$ or $a \geq 6.5$, 
period-3 attractor disappears, and in this case there are total three attractors, positive infinity and negative infinity besides the period-1 attractor. When $a\leq 2.4$, the basin of attraction of 
the period-1 attractor occupies the central white region in Figure \ref{FBB}. When $a>2.4$, three leaves 
appears within the basin of attraction of the period-1 attractor, and they form the basin of attraction of the period-3 attractor (Figure \ref{POA3}). As $a$ increases, the basins of attraction of all four attractors 
intertwine into the three leaves, and fractal basin boundaries are formed (Fig. \ref{FBB}). The 
fractal basin boundaries offer a new kind of sensitive  dependence on initial condition. When the 
market reaches the three leaf regime, the final attractor is very sensitive to its initial condition. A 
small perturbation of the initial condition can perturb into all possible attractors! For instance, 
one initial condition on one of the three leaves leads to the period-1 attractor (so the market will 
reach near the market equilibrium), a small perturbation of the initial condition can leads to 
the period-3 attractor (the market will enter recession and large growth cycle) or one of the infinity attractors (the market will experience irrational exuberance and flash crash). So the three leaves are really 
the market ``danger zone". When $a \geq 6.5$, period-3 ceases to an attractor and its basin of attraction disappears (Figure \ref{FCE}), but the three leaf region still has the fractal basin boundaries of the remaining three attractors --- period-1, positive infinity and negative infinity. So when the fluctuation of the determinants of demand and supply is strong ($a$ is large enough), a danger zone (the three leaves) always exists!

\section{Conclusion}

We introduced a dynamical system model on the dynamics of demand and supply via generalizing the Marshall model to incorporate collectability and saturation factors. Collectability and saturation happen more often than one thought, for instance, many stocks are over-valued (collectability) and under-valued (saturation). Under the 
Marshall model, the dynamics of demand and supply has one global attractor (the market equilibrium). Incorporating the collectability and saturation factors, the dynamics of demand and supply has as many as 
four attractors representing the market regular fluctuation near the market equilibrium, recession (depression) and large growth cycle, and irrational exuberance and flash crash. So our model captured more market phenomena. Our model revealed a ``danger zone" where fractal basin boundaries exist. When the market enters the danger zone, small perturbations can lead to all four attractors, i.e. small perturbations can cause the market 
to experience fluctuation near the market equilibrium, recession (depression) and large growth cycle, and irrational exuberance or flash crash.

\end{document}